\documentclass[12pt,preprint]{emulateapj}
\begin{document}

\parindent=1.0cm

\title{The Stellar Contents of Intermediate Mass Disk Galaxies in the Virgo Cluster. 
I. GMOS Spectra \altaffilmark{1} \altaffilmark{2} \altaffilmark{3}}

\author{T. J. Davidge}

\affil{Dominion Astrophysical Observatory,
\\National Research Council of Canada, 5071 West Saanich Road,
\\Victoria, BC Canada V9E 2E7\\tim.davidge@nrc.ca; tdavidge1450@gmail.com}

\altaffiltext{1}{Based on observations obtained at the Gemini Observatory, which is
operated by the Association of Universities for Research in Astronomy, Inc., under a
cooperative agreement with the NSF on behalf of the Gemini partnership: the National
Science Foundation (United States), the National Research Council (Canada), CONICYT
(Chile), Minist\'{e}rio da Ci\^{e}ncia,
Tecnologia e Inova\c{c}\~{a}o (Brazil) and Ministerio de Ciencia, Tecnolog\'{i}a e
Innovaci\'{o}n Productiva (Argentina).}

\altaffiltext{2}{This research has made use of the NASA/IPAC Infrared Science Archive,
which is operated by the Jet Propulsion Laboratory, California Institute of Technology,
under contract with the National Aeronautics and Space Administration.}

\altaffiltext{3}{This research makes use of the Virgo cluster galaxy database 
of McDonald et al. (2011).}

\begin{abstract}

	The stellar contents of six intermediate mass early-type disk galaxies in 
the Virgo cluster are examined using deep long slit spectra. The isophotal 
and photometric properties of the galaxies at near- and mid-infrared wavelengths 
are also examined. Characteristic ages and metallicities are found by making 
comparisons with the strengths of lines measured 
from model spectra. The light from the central regions of these 
galaxies at visible/red wavelengths is not dominated by old populations. Rather, 
the central regions of four galaxies (NGC 4305, NGC 4306, NGC 4497, and NGC 4620) 
are dominated by populations with ages $\sim 1.5 - 3$ Gyr. Centrally-concentrated 
line emission is found in two of the galaxies (NGC 4491 and NGC 4584), 
and the relative strengths of H$\alpha$ and [SII]6746 are consistent with 
this emission originating in star-forming regions. 
The disks of these galaxies are dominated by populations that are at least 1 Gyr older 
than those near the centers, indicating that the quenching of star formation progressed 
from large radii inwards, and did not occur over a short timescale. 
NGC 4497 has the oldest disk, with a luminosity-weighted 
age of 10 Gyr. The metallicities of the galaxies in this sample 
are consistent with their integrated brightnesses, suggesting that 
they have not been subjected to large-scale stellar stripping. 
[Mg/Fe] is roughly solar, suggesting that these systems retained and enriched 
gas over time scales of at least 1 Gyr. The progenitors of these galaxies 
were likely late-type galaxies that were accreted by Virgo during 
intermediate or early epochs, and have since been depleted of gas and dust.

\end{abstract}

\keywords{galaxies:stellar contents -- galaxies:evolution}

\section{INTRODUCTION}

	The Virgo cluster is the nearest large concentration of galaxies, and so 
is an important laboratory for exploring galaxy evolution in dense environments.
Understanding the stellar content of Virgo galaxies is fundamental to developing a 
picture of how the cluster formed and evolved. The analysis of integrated 
light is the primary means of studying the stellar contents 
of galaxies and star clusters in Virgo, as even the most luminous 
stars can only be resolved in the lowest density regions of its member galaxies. 
Spectroscopic observations at visible wavelengths are 
of particular importance for examining stellar content 
as there are numerous well-calibrated atomic and molecular transitions at these 
wavelengths (e.g. Worthey et al. 1994). For example, features at visible wavelengths 
such as H$\beta$ and the G-band ease the age-metallicity degeneracy (e.g. Worthey 
1994). However, spectra at visible wavelengths do not sample 
the full range of stellar content. While stars near the MSTO 
in old and intermediate age systems may contribute significantly 
to integrated light at visible wavelengths, the contribution from
stars that have effective temperatures (T$_{eff}$) $> 10^4$ K, such as young MS 
stars, BHB and PAGB stars, or from stars that have T$_{eff} <$ a few 
times $10^3$ K (e.g. RGB-tip and AGB stars) may be modest. Extending the wavelength 
coverage can thus contribute additional information for probing 
stellar content in Virgo galaxies, such as finding 
evidence for very low levels of recent star formation that may be 
missed at visible wavelengths (e.g. Vazdekis et al. 2016). Obscuration by dust will 
also complicate efforts to examine integrated light at visible wavelengths.

	This is the first of two papers that examines the spectra 
of six intermediate mass disk galaxies in the Virgo cluster. Long slit spectra 
covering wavelengths from $0.47\mu$m to $2.3\mu$m and that sample angular offsets 
out to 15 arcsec from the galaxy centers have been obtained 
for these galaxies. This wavelength interval contains a mix of classic stellar 
population diagnostics at blue/visible wavelengths, such as H$\beta$ and lines of Mg 
and Fe. It also samples spectroscopic signatures of the reddest evolved stars that 
contribute a significant amount of the light at red and near-infrared (NIR) 
wavelengths, such as the Ca triplet near $0.86\mu$m and the first and second 
overtone CO bands between 1.5 and $2.4\mu$m. Signatures of C stars 
may also be detected at the long wavelength end of these observations. 
Even though the rapid pace of evolution on the upper AGB makes C stars relatively rare, 
spectroscopic signatures of C stars have been found in the integrated 
light from star clusters (Mouhcine et al. 2002; Davidge 2018) and nearby galaxies 
(Miner et al. 2011; Davidge et al. 2015; Davidge 2016). 

	All six galaxies are of type dE(di), as defined 
by Lisker et al. (2006a). Galaxies of this type show evidence of a disk in the 
form of structures like spiral arms, bars, lenses, and/or a flattening of 
the light profile that is consistent with rotational support. dE(di)s account for 
$\sim 10\%$ of the dE population in Virgo, and tend to inhabit the bright end of the 
dE luminosity function, where they account for $\sim 50\%$ of all objects 
(Lisker et al. 2006a). That dE(di)s populate the bright end of the dE LF may indicate 
that there is a minimum mass threshold to sustain organized disk structures like bars 
and spiral arms (Lisker et al. 2008). Many properties of dE(di)s differ from those of 
`classical' dEs, suggesting that they are a distinct class of 
objects (Lisker er al. 2006a). 

	Five of the galaxies (NGC 4305, NGC 4491, 
NGC 4497, NGC 4584, and NGC 4620) form the sub-sample of 
dE(di) galaxies classified as `dwarfish S0/Sa' by Lisker et al. (2006a). 
These are among the brightest dE(di)s, and 
lack the large bulge seen in more massive classical early-type disk galaxies. 
Galaxies of this nature may serve as a bridge 
between fainter dE(di) galaxies and larger disk systems. 
Various properties of the target galaxies are listed in Table 1, 
and the absolute $K$ brightnesses of these galaxies 
are similar to those of the LMC and M33 (M$_K \sim -20.4$). 

	The sixth galaxy is NGC 4306, which is close to NGC 4305 on the 
sky and could be observed at the same time as NGC 4305 as the GMOS slit can sample 
both systems at once. The close proximity of NGC 4306 to NGC 4305 is fortuitous, 
as NGC 4306 is one of only a handful of Virgo dEs that Lisker et al. (2006a) identify 
as having a `certain disk', and kinematic measurements made by Toloba et al. (2011) 
confirm that NGC 4306 is rotationally supported. NGC 4306 thus is a bona fide dE(di), 
having a morphological kinship with the other five galaxies examined in this 
paper. As such, NGC 4306 serves as a link to fainter dE(di)s.

\begin{deluxetable*}{cccccccc}
\tablecaption{Galaxy Properties}
\startdata
\tableline\tableline
NGC \# & VCC \# & $r'$\tablenotemark{a} & $K$\tablenotemark{b} & M$_K$\tablenotemark{c} & r$_e$ \tablenotemark{d} & $\Delta \alpha$ \tablenotemark{e} & Morphology \tablenotemark{f} \\
 & & (mag) & (mag) & (mag) & (arcsec) & (arcmin) & \\
\hline
4305 & 0522 & 12.7 & 10.2 & --20.9 & 29.0 & 133.5 & Sa \\
4306 & 0523 & 13.1 & 10.9 & --20.2 & 16.9 & 130.3 & d:SB0(s),N \\
4491 & 1326 & 12.6 & 9.8 & --21.3 & 25.3 & 54.5 & SBa(s) \\
4497 & 1368 & 12.5 & 9.9 & --21.2 & 35.2 & 47.2 & SB0(s)/SBa \\
4584 & 1757 & 13.1 & 10.9 & --20.2 & 19.6 & 117.8 & Sa(s) pec \\
4620 & 1902 & 12.6 & 10.4 & --20.7 & 18.2 & 166.8 & S0/Sa \\
\tableline
\enddata
\tablenotetext{a}{$r'$ brightness from SDSS data release 6 (Adelman-McCarthy et al. 2008).}
\tablenotetext{b}{From 2MASS Extended Object catalogue, 
except for NGC 4491, where $K$ is from the 2MASS Large Galaxy Atlas 
(Jarrett et al. 2003).}
\tablenotetext{c}{Based on a distance modulus of 31.3 (Mei et al. 2007).}
\tablenotetext{d}{Half light radius in arcsec from McDonald et al. (2011).}
\tablenotetext{e}{Angular offset from M87 in arc minutes.}
\tablenotetext{f}{From Binggeli et al. (1985).}
\end{deluxetable*}

	All six galaxies have red colors, with $g'-r' \sim 0.7$ 
and $i'-z' \sim 0.2$. To the extent that dust does not define the photometric properties
of Virgo galaxies (e.g. Roediger et al. 2011) then these galaxies might then be 
expected to harbor large intermediate age and old populations, with negligible 
contributions to the light from recent star formation. While 
this is shown to be the case for four of the galaxies, 
recent star formation dominates the light from the central regions 
of two others. The visible/red colors of these systems are thus not faithful 
indicators of age, and this is likely due to dust.

	In the present paper spectra of the galaxies obtained with the 
Gemini Multi-Object Spectrographs (GMOSs) on Gemini North (GN) and Gemini South (GS) 
are discussed. Spectra from GS that span the $0.47 - 1.1\mu$m wavelength interval 
were recorded for all six galaxies. Two galaxies (NGC 4305 and NGC 4620) were 
also observed from GN, and those spectra extend to $0.39\mu$m. 
The spectra sample the galaxy nuclei and the surrounding disk, 
allowing stellar content to be examined over 
a range of radii. Luminosity-weighted ages and metallicities are 
estimated through comparisons with model spectra.
Spectra of these galaxies at longer wavelengths obtained with the Flamingos-2 
spectrograph on GS will be the subject of a separate paper (Davidge 2018, 
in preparation).

	The spectra are used to estimate the ages and metallicities 
of these systems, including a search for radial trends within the galaxies. The age 
of these systems is of particular interest for probing their origins -- if their 
light is dominated by a uniformly old population then they are probably 
long-time residents of the Virgo cluster. The use of spectra allows ages to be 
estimated that are free of the age-metallicity degeneracy that 
plagues estimates made from visible broad-band photometry alone. Spectra can also 
be used to identify young stellar components that may be reddened by dust, 
thereby making their detection difficult with broad-band photometry alone. 
Finally, the radial colors of dE(di)s suggest that there are radial population 
gradients (Lisker et al. 2008). Age gradients in the galaxies might 
provide insights into the timescale of the removal of the interstellar 
mediums from the galaxies. The absence of an age gradient would indicate that 
star formation was shut down throughout the galaxy over a narrow time frame, 
whereas a large age gradient would indicate that the quenching of star formation was 
the result of a process that affected some parts of the galaxy more quickly 
than others.

	Three of the galaxies (NGC 4305, NGC 4491, and NGC 4584) are in the 
Herschel Reference Survey (HRS; Boselli et al. 2010), and there are HRS spectra 
for these galaxies that cover wavelengths from 0.39 to 0.70$\mu$m. 
However, HRS spectra were obtained via drift-scanning, 
and so do not contain spatial information. While spectra of this nature 
are useful for assessing the stellar content of a galaxy in its entirety, it 
will be shown later in this paper that all three galaxies 
contain population gradients, and two of these (NGC 4491 and NGC 4584) 
have a prominent centrally-concentrated young component.

	Details of the observations and a description of the processing that 
was applied to extract the spectra are presented in Section 2. The photometric and 
isophotal properties of the galaxies provide supplemental insights into their 
past histories, and measurements made from archival NIR and 
mid-infrared (MIR) wavelengths are discussed in Section 3. The characteristics of the 
spectra, including line strength measurements and radial variations in spectroscopic 
properties, are the subject of Sections 4 and 5. Comparisons are made with 
model spectra in Section 6, where luminosity-weighted ages and metallicities are 
estimated. A summary and discussion of the results follows in Section 7.

\section{OBSERVATIONS AND REDUCTION PROCEDURES}

\subsection{GMOS-S Spectra}

	Spectra of all six galaxies were recorded with GMOS (Hook et al. 2004) 
on GS as part of program GS-2017A-Q-81 (PI: Davidge). The detector 
in GMOS-S is a mosaic of three Hamamatsu CCDs, 
the general characteristics of which are discussed by Hardy et al. (2012).
Each CCD is made up of $15\mu$m square pixels in a 2048 $\times$ 4176 pixel 
format, with an on sky angular sampling of 0.08 arcsec pixel$^{-1}$. The Hamamatsu CCDs 
have a higher quantum efficiency over a broader wavelength range than 
the EEV CCDs that made up the previous GMOS detector, and are also 
less susceptible to fringing. 

	Light was dispersed with the R150 grating (150 l/mm; $\lambda_{blaze} 
= 0.72\mu$m), with a GG451 filter deployed to block signal from higher orders. 
While second order light is then present at wavelengths longward 
of $0.9\mu$m, contamination is likely to be minor given the red colors of the 
galaxies and the dispersed nature of the second order light. A 2 arcsec wide slit 
was used for all observations, and the spectral resolution of a source that uniformly 
fills the slit is $\Delta \lambda/\lambda = 315$ at the blaze wavelength. 
Spectra were recorded with two wavelengths at the center of the detector array 
($0.65\mu$m and $0.70\mu$m) so that breaks in spectral 
coverage caused by gaps between CCDs could be filled.
The detectors were binned $2 \times 2$ during readout. 
The dates when galaxies were observed are listed in Table 2, while the 
orientation of the slit for each galaxy is shown in Figure 1. Aside from 
NGC 4305 and NGC 4306, the slit orientations were selected based on the availability 
of guide stars that could be acquired by the on-instrument wavefront sensor in GMOS, 
and the need to sample the galaxies in a manner that ensured clean sky subtraction 
when stepping along the slit (see below). 

\begin{deluxetable}{cccc}
\tablecaption{Dates of Observation}
\startdata
\tableline\tableline
Telescope & Galaxy & Date (2017, UT) & Grating \\
\tableline
GS & NGC 4305 & April 13 & R150 \\
 & NGC 4606 & April 13 & R150 \\
 & NGC 4491 & April 24 & R150 \\
 & NGC 4497 & April 23 & R150 \\
 & NGC 4584 & April 13 & R150 \\
 & NGC 4620 & April 13 & R150 \\
 & & & \\
GN & NGC 4305 & May 27 & B600 \\
 & ... & June 26 & B600 \\
 & ... & June 26 & R400 \\
 & & & \\
 & NGC 4620 & June 27 & B600 \\
 & ... & June 28 & R400 \\
\tableline
\enddata
\end{deluxetable}

\begin{figure*}
\figurenum{1}
\epsscale{1.0}
\plotone{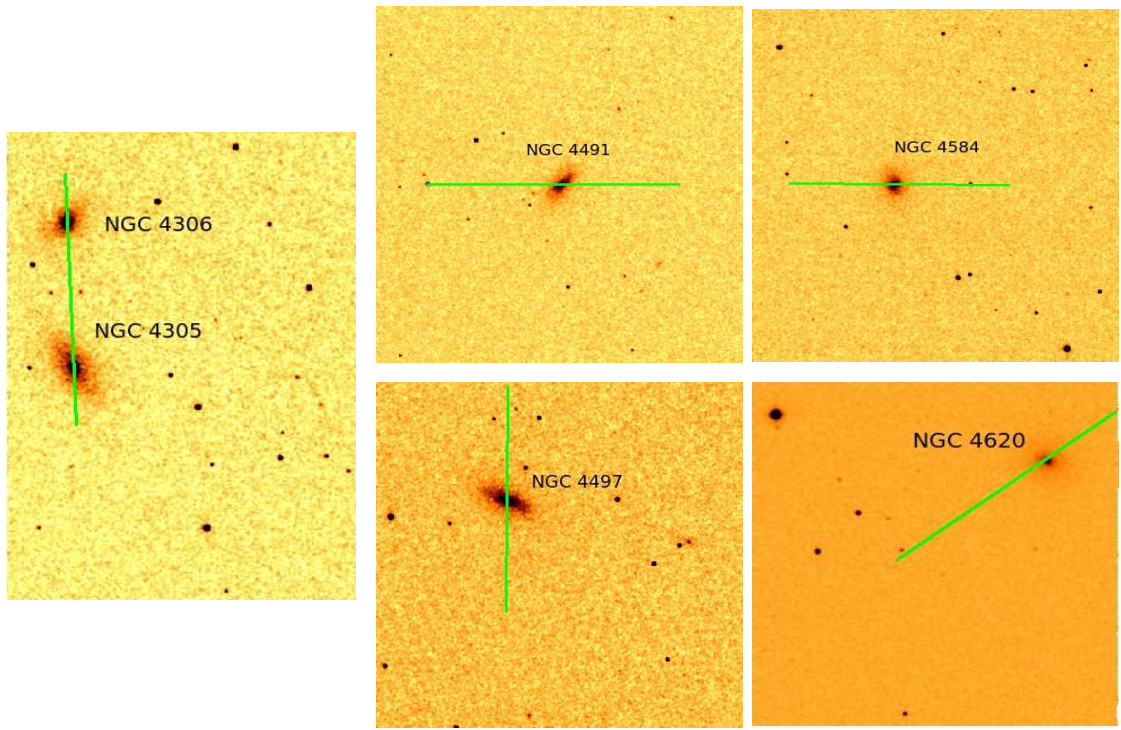}
\caption{GMOS slit orientations, as shown on $J-$band 2MASS images. 
The slit in each panel has a length of 5 arcmin. North is at the top, and East 
is to the left.}
\end{figure*}

	The suppression of the numerous bright telluric emission bands 
at wavelengths longward of $0.7\mu$m can be a challenge for observations 
of faint, extended objects. To facilitate sky subtraction, NGC 4491, 4497, 4584, 
and 4620 were nodded between two locations on the 5.5 arcmin slit at each 
wavelength setting. Each nodded pair of observations could then be subtracted, 
thus providing an efficient means of removing sky emission lines. Interference 
fringes are also suppressed. Two 300 sec exposures were obtained at each 
pointing and wavelength setting, with the result that eight 300 sec exposures were 
recorded for each galaxy.

	The angular separation between NGC 4305 and NGC 4306 
is such that both can be observed with GMOS at the same time, and this was done 
here. The total exposure time is the same as for the other galaxies, 
with four 300 sec exposures recorded at each wavelength setting.
While saving time, the separation between the galaxies is such that 
on-slit nodding is not possible. While the sky level can be 
measured in the space between the two galaxies, uncertainties in the 
flat-field correction and optical distortions within GMOS are such that 
the suppression of sky lines proved not to be as good as for the other 
galaxies. This limits the radial extent of the NGC 
4306 spectra, which has the faintest surface brightnesses at 
a given angular offset from the galaxy center in the six galaxy sample. 
Comparisons with nodded spectra recorded from GN (Section 2.2) indicate 
that line strengths in the NGC 4305 GS spectra are not affected by sky line 
residuals.

	Calibration data were also recorded. These 
include (1) spectra of an early-type white dwarf (EG 81) 
to monitor telluric absorption features and the wavelength response of the 
atmosphere, telescope, and instrument; (2) flat-field frames, constructed from 
observations of dispersed light from a continuum source in 
the Gemini Facility Calibration Unit (GCAL); (3) exposures 
that monitor the detector bias level; and (4) spectra 
of the GCAL CuAr arc lamp. Following standard observing procedures for 
queue-scheduled GMOS observations, the flat field frames 
were recorded midway through the observing sequence for each galaxy, while 
the bias and arc exposures were recorded on the days following nights 
when program targets were observed.

\subsection{GMOS-N Spectra}

	Spectra of NGC 4305 and NGC 4620 were recorded with GMOS on 
GN as part of program GN-2017A-Q-78 (PI Davidge). The detector in GMOS-N 
is a mosaic of Hamamatsu CCDs that is similar to its GMOS-S counterpart. 
The GN data were recorded with $2 \times 2$ pixel binning during read-out.

	The GN spectra have a higher spectral resolution than the GS spectra. 
Two sets of spectra were recorded for each galaxy: 
one with the B600 (600 l/mm; $\lambda_{blaze} = 0.461\mu$m) grating, and 
the other with the R400 (400 l/mm; $\lambda_{blaze} = 0.764\mu$m) grating.
The B600 spectra were recorded with central wavelengths of 0.54 and 
$0.545\mu$m, while the R400 spectra have central wavelengths of 0.75 and $0.80\mu$m. 
Together these spectra cover wavelengths from 
0.38 to $1.1\mu$m. The spectra were recorded using a 1 arcsec slit, with a GG455 
filter deployed for the R400 observations to block light from higher orders. 

	It was decided not to observe NGC 4306 with NGC 4305, making it possible 
to nod NGC 4305 along the slit to facilitate the removal of telluric emission lines. 
The processed GN spectra of NGC 4305 then provide a means of gauging if there are 
systematic effects in sky subtraction in the GS spectra of this galaxy. 
The spectra of NGC 4620 were also recorded with on-slit nodding. Calibration 
images that parallel those obtained on GS were also recorded. The DA white 
dwarf EG 131 was observed to monitor telluric absorption features and system throughput.

	Two 900 second exposures were recorded at each central wavelength for NGC 
4305, while four 600 sec exposures were recorded of NGC 4620. The total exposure 
times are then 3600 sec for NGC 4305 and 4800 sec for NGC 4620. 
The dates when the GN spectra were recorded are listed in Table 2. 
The observations of NGC 4305 with the B600 grating were recorded over two nights.

\subsection{Reduction}

	The processing of the spectra was carried out in two stages, and the 
various steps in each stage are described below.

\subsubsection{Removal of instrument signatures and the sky}

	To begin, the overscan level of each CCD was measured and subtracted on an 
exposure-by-exposure basis, and a nightly bias frame was subtracted from the results. 
Cosmetic defects and cosmic rays were repaired through interpolation. 
Exposures of each galaxy at the same nod position and with the same central 
wavelength were then averaged together, and the results were divided by 
a processed flat-field frame. 

	Geometric distortions produced by the GMOS optics 
bow the signal along the spatial direction, distorting spectra of extended objects. 
Distortion maps were constructed by tracing emission 
lines perpendicular to the dispersion direction in the arc spectra, and using the 
results to construct warping functions that linearized these features. 
With the exception of the GS observations of NGC 4305 and NGC 4306 (see below), 
the sky was then removed by differencing spectra of each galaxy recorded at 
the same central wavelength but at different nod locations. 
The sky line residuals are small in the reduced data (Section 4), 
indicating that the sky brightness was stable on time scales in excess of the many 
hundreds of seconds during which nodded pairs were recorded. 

	A different sky subtraction procedure was applied to the GS observations of 
NGC 4305 and NGC 4306, as these data were recorded without slit nodding (Section 
2.1). The sky level for these observations was measured by taking the mean signal 
midway between the galaxies and near the ends of the slit. The results were then 
subtracted on a column-by-column basis. The residuals at wavelengths near sky emission 
lines are larger than those in spectra in which nodding was employed, and 
this is most noticeable in the low surface brightness outer 
regions of NGC 4306. The larger residuals are a 
consequence of modest ($\sim \pm 1$\%) uncertainties in flat-fielding and 
the optical distortion corrections, both of which 
are negated when nodded spectra are subtracted.

\subsubsection{Extraction of spectra, wavelength calibration, and correction 
of throughput variations.}

	Mean spectra covering various radial intervals were extracted from 
the sky-subtracted two-dimensional spectra of each galaxy. 
A central spectrum (Region 1) was extracted within $\pm 0.55$ arcsec of the 
nucleus to more-or-less sample the angular resolution defined by the 
seeing, and progressively wider radial gathers (Regions 2 -- 5) were 
defined to maintain a roughly constant galaxy light level in each interval. 
Spectra on both sides of the nucleus were combined, and 
the radial extraction intervals are listed in Table 3. 
Efforts to conserve total signal notwithstanding, the S/N ratio of the 
extracted spectra drops as progressively larger angular offsets 
from the galaxy center are sampled, since the contribution from sky noise grows 
when spectra are combined over wider and wider angular intervals in the 
low surface brightness outer regions of a galaxy. 

\begin{deluxetable}{ccc}
\tablecaption{Radial extraction limits}
\startdata
\tableline\tableline
Region \# & Radial extent & Projected spatial extent\tablenotemark{a}\\
 & (arcsec) & (pc) \\
\tableline
1 & 0 -- 0.55 & 0 -- 50 \\
2 & 0.55 -- 1.75 & 50 -- 150 \\
3 & 1.75 -- 3.90 & 150 -- 340 \\
4 & 3.90 -- 8.40 & 340 -- 740 \\
5 & 8.40 - 15.90 & 740 -- 1400 \\
\tableline
\enddata
\tablenotetext{a}{Based on a distance modulus of 31.3 (Mei et al. 2007).}
\end{deluxetable}

	The extracted spectra were wavelength calibrated by applying a 
dispersion solution obtained from the CuAr arcs.
The spectral response at this point in the processing contains contributions 
from the system optics (telescope $+$ instrument), the detector, 
the atmosphere, and the flat-field source. 
These were suppressed by ratioing the extracted spectra with the 
spectrum of a reference star that had been processed using the same procedures 
as those applied to the galaxy data. A low-order continuum 
function was then fit to and divided out of each spectrum. Doppler-corrections were 
applied to place the spectra in the rest frame. 

	It should be noted that Gemini's baseline calibration policy is to record 
one reference star spectrum for each science program. While this typically 
works well for suppressing O$_2$ absorption at 0.69 and $0.76\mu$m, 
there can be issues suppressing H$_2$O bands near $0.72\mu$m, $0.81\mu$m 
and $0.93\mu$m. These H$_2$O bands are not saturated, and their depths vary with 
atmospheric water content. GN is typically very dry, and the night-to-night 
variation in the depths of the H$_2$O bands is usually modest. However, GS is at a 
lower altitude than GN, and night-to-night variations in H$_2$O content can be 
significant. Four of the galaxies observed from GS were observed on the same night 
(April 13) as the reference star, and the H$_2$O bands are suppressed in the reduced 
spectra. However, NGC 4491 and NGC 4497 were not observed on the same night as the 
reference star, and deep residual H$_2$O absorption between 0.9 and $1.0\mu$m 
remained in the NGC 4497 spectrum after division by the reference star. 
To correct for this, a template was constructed that forced the wavelength 
intervals containing H$_2$O in the NGC 4497 spectrum to match the mean of the 
other five galaxies at these wavelengths, but that did not 
alter individual atomic absorption or emission features. There 
are CN and TiO band heads near $0.9\mu$m, and this procedure forces 
the depths of these bands in NGC 4497 to match those in the other galaxies.

\section{ISOPHOTAL PROPERTIES}

	The isophotal properties of galaxies form part of the fossil record 
that can be mined to explore their evolution. In the current study, the isophotal 
characteristics of the six galaxies are examined using images from the 2MASS and WISE 
All Sky surveys. McDonald et al. (2011) examined the light profiles of these galaxies, 
fitting models that include a nucleus, a Sersic bulge, and a disk. The Sersic 
indices measured for the central regions of NGC 4305, NGC 4306, NGC 4491, and NGC 4497 
are consistent with those found in late-type spirals and dEs 
(Table 2 of McDonald et al. 2011). However, the Sersic indices for 
the centers of NGC 4584 and NGC 4620 are consistent with those found in Sa and Sb 
galaxies, suggesting that classical bulges may be present.

	We do not attempt to duplicate the McDonald et al. (2011) structural 
analysis. Rather, the current discussion is restricted to isophote shape and 
color profiles. The isophotal measurements were made with the STSDAS routine 
$ellipse$, which applies the iterative method described by Jedrzejewski (1987).
With the exception of extreme cases such as dust-enshrouded starburst systems, 
the light in $J$ and $K$ traces stellar mass, and is less 
susceptible to stochastic variations that influence isophotes 
at shorter wavelengths when very young populations are present. 
The angular resolution of the 2MASS data measured 
directly from stellar profiles in the images is $\sim 3$ arcsec, which is 
broader than the $\sim 1$ arcsec FWHM resolution of the GMOS observations. 
Measurements were also made from WISE All Sky Survey (Wright et al. 2010) W2 
($\lambda_{cen} \sim 4.6\mu$m) images. Light in this filter is of 
potential interest for determining if there is a contribution from hot 
(logT$_{eff} \geq$ 2.0) dust emission, such as might originate in circumstellar 
envelopes, young star-forming regions, and/or near AGN. 

	Bars are seen in some massive lenticulars (e.g. Dullo 
et al. 2016), and similar structures will complicate the isophotal analysis if present 
in these galaxies. While some of the galaxies are classified as barred, the residual 
images produced by subtracting the isophotal fit from the 
original images do not contain obvious large-scale residuals, 
indicating that the isophotal fits have faithfully tracked these structures. 
The angular resolution of the 2MASS and WISE images is such that bars 
with angular sizes $\leq 1$ arcsec will go undetected.

	The $J-K$ profiles obtained from the 2MASS images are shown in Figure 2. 
Residual images suggest that the jitter in the $J-K$ profiles 
is likely not a result of spiral structure, but may instead be due to noise. 
The central $J-K$ colors of the galaxies differ by $\sim \pm 0.1$ magnitudes, 
suggesting that differences in the central spectroscopic 
properties of the galaxies might be expected. NGC 4491 has the reddest central $J-K$ 
($J-K \sim 1.1$), while NGC 4306 and NGC 4497 are at the other extreme, with 
$J-K \sim 0.85$. McDonald et al. (2011) determined the half light radius of bulge 
light for each galaxy, and the result is marked in the top row of Figure 2. 
The bulge radius coincides with features in the $J-K$ color profiles of 
NGC 4305, NGC 4497, and NGC 4584.

\begin{figure*}
\figurenum{2}
\epsscale{1.0}
\plotone{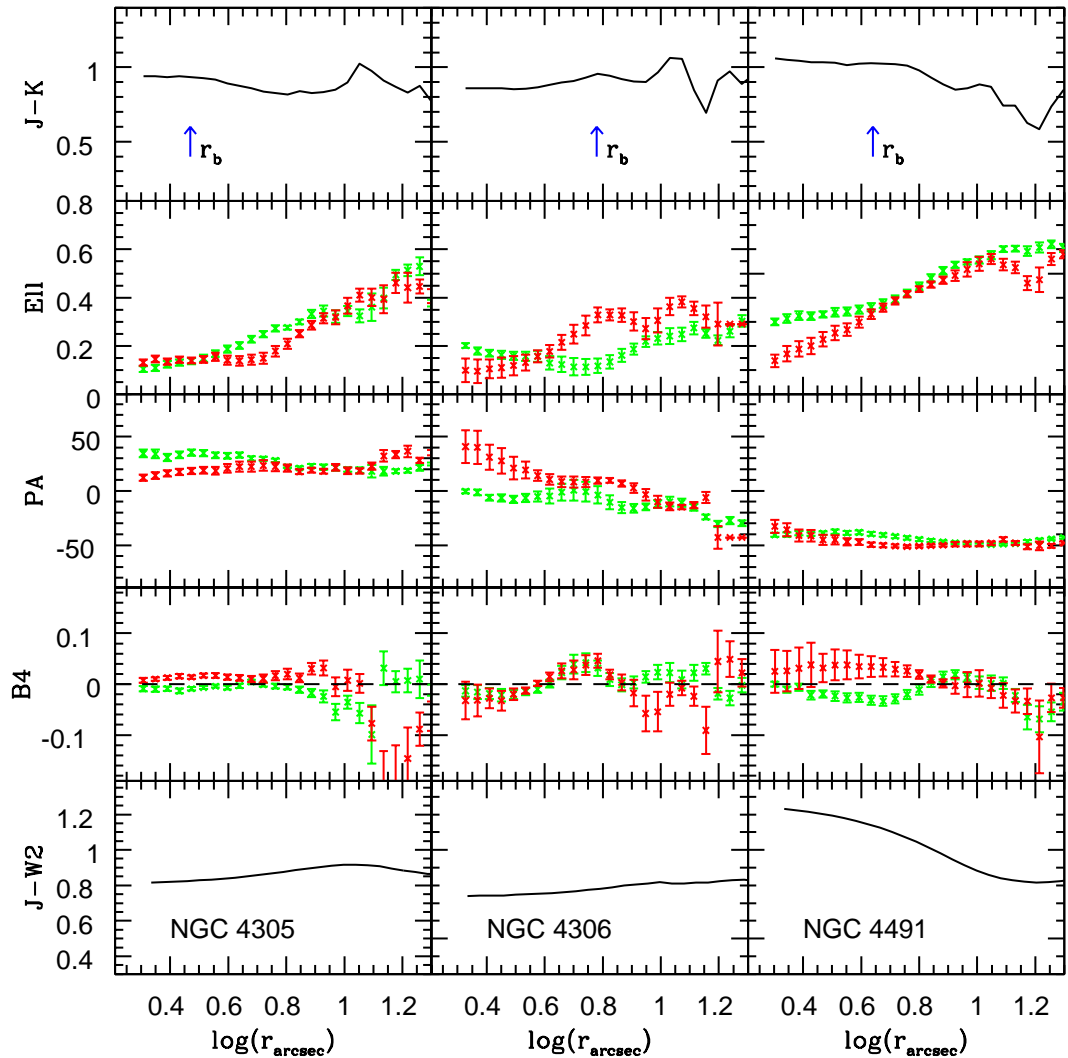}
\caption{Isophotal properties obtained from 2MASS and WISE images. 
The $J-K$ color, ellipticity (ELL), position angle (PA), 
4th order cosine coefficient (B4), and $J-W2$ color are shown for each galaxy. 
Angular distances are measured along the major axis, and the 
effective radius of the bulge in each galaxy calculated by McDonald et al. (2011) 
is indicated in the top panels. Measurements made from the $J$ images are 
shown in green, while those from the $K$ images are in red. 
The error bars are the uncertainties calculated by $ellipse$. 
The $J-W2$ colors were computed using 2MASS $J$ 
images that had been smoothed to match the FWHM of the W2 data.}
\end{figure*}

\begin{figure*}
\figurenum{2}
\epsscale{1.0}
\plotone{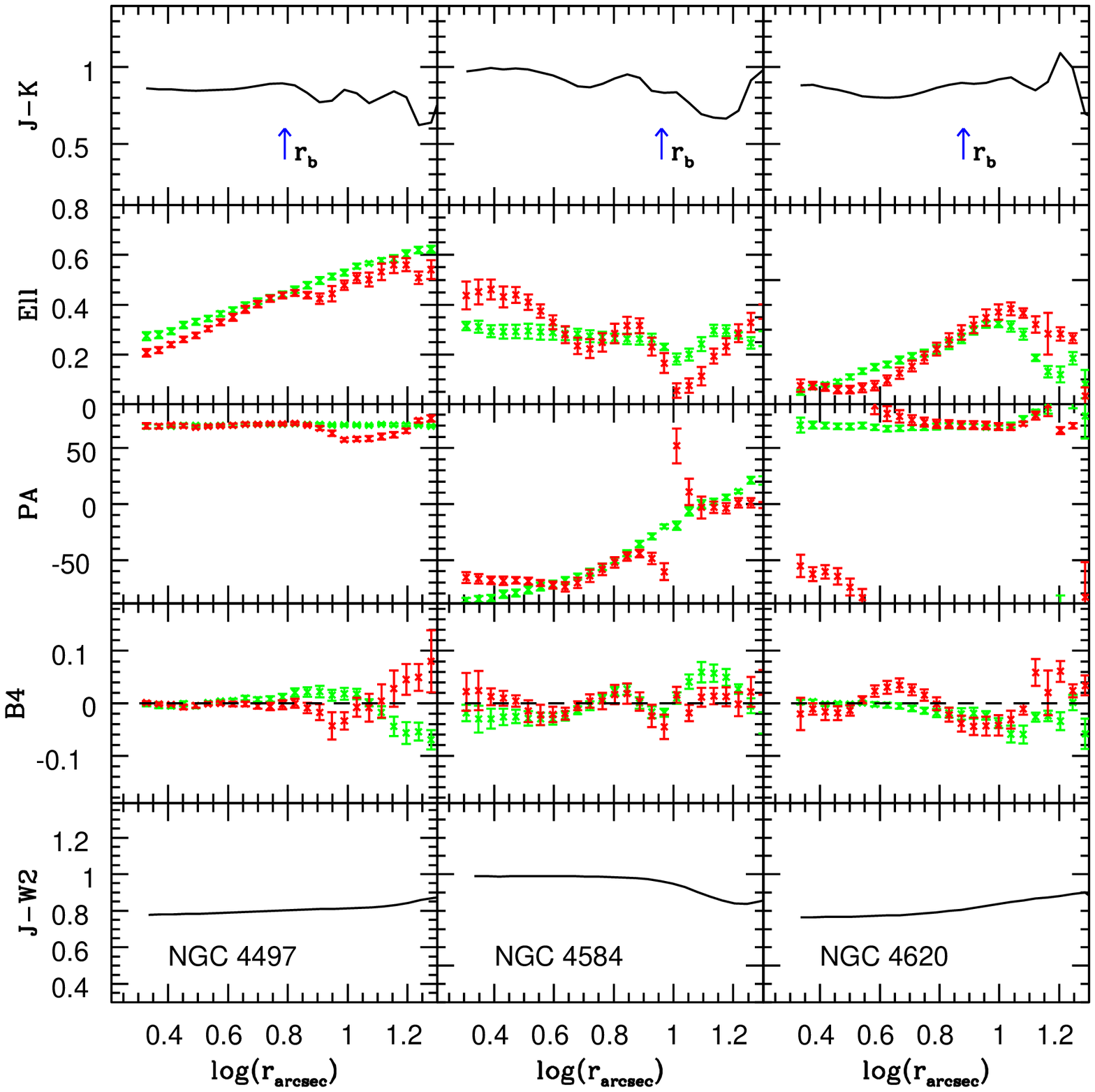}
\caption{Figure 2, con't}
\end{figure*}

	Bulges that are redder than the surrounding 
disk are a common feature in larger S0s (e.g. Xiao et al. 2016). 
Neglecting the color measurements in the outermost bins, which 
are susceptible to errors in sky subtraction and noise, 
then there are systematic $J-K$ gradients in NGC 4305, NGC 4491, and NGC 4584, 
in the sense that $J-K$ becomes smaller towards larger radii. 
There may also be a shallow gradient in NGC 4497. In contrast, 
NGC 4306 and NGC 4620 appear to have a more-or-less constant $J-K$ color.

	In the absence of dust obscuration, $J-K$ gradients in intermediate 
age and older systems are typically interpreted as being due to 
metallicity-driven changes to the effective temperatures of giants. 
Radial changes in the strengths of lines that trace metallicity 
might then be expected in the spectra of NGC 4305, NGC 4491, NGC 4497, and NGC 4584. 
In contrast, the radial spectroscopic properties of NGC 4306 and NGC 4620 might 
be expected not to change with radius given the $J-K$ profiles of these galaxies. 
These predictions are broadly consistent with the trends in the spectra that are 
discussed in Section 5.

	McDonald et al. (2011) estimate the ratio of 
bulge to total light (B/T) in $g$ and $H$, and these measurements 
provide additional clues into the radial distribution of stellar 
content in these galaxies. The B/T ratios in $H$ span a large range, amounting to 
only a few percent for NGC 4305, NGC 4491, and NGC 4497, while B/T = 0.10 for NGC 
4306 and 0.20 for NGC 4584 and NGC 4620. McDonald et al. (2011) also 
find that B/T in $g$ can be up to a factor of ten larger than in $H$. This 
difference between B/T in $g$ and $H$ suggests that there is 
a larger fractional contribution from blue stars to 
the visible light from the central regions of these galaxies than in the 
surroundings. This is additional evidence that stellar content 
varies with radius in some of these systems, and this is confirmed with the spectra.  

	The ellipticity and position angle profiles in Figure 2 
point to a mix of structural characteristics, and NGC 4584 
stands out in two ways. First, the isophote position 
angle stays more-or-less constant in most of the galaxies, indicating 
that there is no large-scale twisting of the isophotes. 
An exception is NGC 4584, where the position angle measurements 
change markedly with radius in both filters. Second, there is a tendency for 
ellipticity to increase with radius in almost all of the galaxies. 
Again, an exception is NGC 4584. Such a radial 
variation in ellipticity could occur if there is a radial change in the 
light contribution made by a (presumably) round nucleus/bulge and the surroundings. 
The galaxies that McDonald et al. (2011) find have the 
smallest B/T ratio also have the steepest ellipticity gradients in Figure 2, 
as would be expected if the ellipticity profile is due to a radial transition 
from a round nucleus to a disk viewed at an oblique angle.

	$ellipse$ performs a Fourier analysis of isophotes. 
The coefficient of the fourth order cosine term -- B4 -- measures 
departures from pure ellipticity (e.g. Carter 1978), that could be attributed to a 
disk, distortions caused by tidal interactions, and/or isolated concentrations of light 
from large star-forming complexes. Isophotes with B4 $> 0$ are `disky', while 
those with B4 $< 0$ are `boxy'. The agreement between the B4 coefficients measured 
in both filters in each galaxy in Figure 2 is far from perfect, 
highlighting the susceptibility of B4 measurements to noise. 
Filter-to-filter disagreements in B4 within the seeing disk (i.e. log(r) $\leq 0.5$) 
are likely due to distortions in the point spread 
function (PSF) that are filter and field specific.

	Given the uncertainties in the B4 measurements, we only consider 
trends in B4 outside of the seeing disk that are detected 
in both filters. For NGC 4305, NGC 4306, NGC 4497, and NGC 4620 
B4 $\sim 0$ within the errorbars. There is no evidence of isophotal twisting 
in the position angles of these galaxies, suggesting that they probably 
have not experienced interactions within the last few disk crossing times. 
However, the situation is different for NGC 4491 and NGC 4584, 
both of which have prominent emission lines in 
their spectra (Section 4). Focusing on radii log(r) $\geq 1.0$ 
then systematic departures of B4 from zero in both bandpasses 
are seen in NGC 4491, where B4 tends to be negative, and NGC 4584 where 
B4 tends to be positive between log(r) = 1 and 1.3. Despite having different 
isophote shapes at large radii, both galaxies show evidence of isophote 
twisting in their position angle profiles.

	The light profiles of bright field stars in the W2 images 
of these galaxies indicate that the angular resolution is $\sim 9$ arcsec FWHM 
\footnote[1]{For comparison, the angular resolution delivered by the satellite optics 
is 6.4 arcsec FWHM for W2 (Wright et al. 2010)}, which is markedly broader than 
the FWHM of stars in the 2MASS images. Examining isophote properties 
on arcsec angular scales in the W2 images is thus problematic, and the W2 images 
are used only to compute $J - W2$ colors. Colors were measured using 
2MASS $J$ images that had been convolved with a Gaussian to match 
the angular resolution of the W2 data. The resulting $J-W2$ profiles are shown in the 
bottom row of Figure 2. 

	For four of the six galaxies (NGC 4305, NGC 4306, NGC 4497, and NGC 4620) 
there is a tendency for $J-W2$ to increase slightly with increasing radius. 
This could occur if the dominant signal in the W2 observations 
at the largest radii shown in Figure 2 is not photospheric in origin. 
There are no polyaromatic hydrocarbon (PAH) emission bands in the wavelength region 
sampled by the W2 filter, and non-photospheric light might then originate in 
circumstellar envelopes associated with a large population of hot stars or a dominant 
intermediate age population at large radii. However, this is unlikely. In Section 
6 it is shown that while the disks are older than the galaxy centers, the central 
regions tend to have ages where the contribution from AGB stars with solar 
metallicities is expected to peak (e.g. Figure 13 of Maraston 2005); 
the amount of thermal emission from circumstellar envelopes might 
then be expected to decrease as one moves to larger radii. The 
slight $J-W2$ gradients in NGC 4305, NGC 4306, NGC 4497, and 
NGC 4620 may be due to subtle systematic errors in the sky background measurements.

	$J-W2$ near the centers of NGC 4491 and NGC 4584 
is much higher than in the outer regions of these galaxies. 
The $J-W2$ colors of young and intermediate age systems 
constructed from the Bressan et al. (2012) isochrones 
are a few tenths of a magnitude smaller than those found near the centers of 
these two galaxies, but agree with what is seen in their 
outer regions, as well as throughout the other four galaxies. 
The red central $J-W2$ colors of NGC 4491 and NGC 4584 are thus 
likely due to thermal emission from hot dust.

	NGC 4491 has the reddest central $J-W2$ color in 
Figure 2. Using Herschel Virgo Cluster Survey 
measurements, Auld et al. (2013) compute a dust temperature of $26.4 \pm 1.4$ K, 
which is the highest temperature of any object in that survey. The presence 
of hot dust in NGC 4491 is thus not surprising. 
In the next section it is shown that the line emission 
in NGC 4491 and NGC 4584 is powered by star formation, and so dust near the 
centers of these galaxies is probably heated by luminous young stars. 

\section{OVERVIEW OF THE CENTRAL SPECTRA}

\subsection{GMOS-S}

	The Region 1 spectrum of each galaxy is shown in Figures 3 - 5. The 
discussion is restricted to the galaxy centers for this 
initial reconnaisance of spectral features, as this is where the S/N ratio is highest. 
Many of the galaxies show gradients in spectroscopic properties, and 
the emission lines that dominate the spectra of NGC 4491 and NGC 4584 disappear at 
the largest angular offsets sampled with GMOS.
The wavelength coverage is divided over multiple figures 
to allow the identification of individual lines and 
facilitate galaxy-to-galaxy comparisons. 

\begin{figure*}
\figurenum{3}
\epsscale{1.0}
\plotone{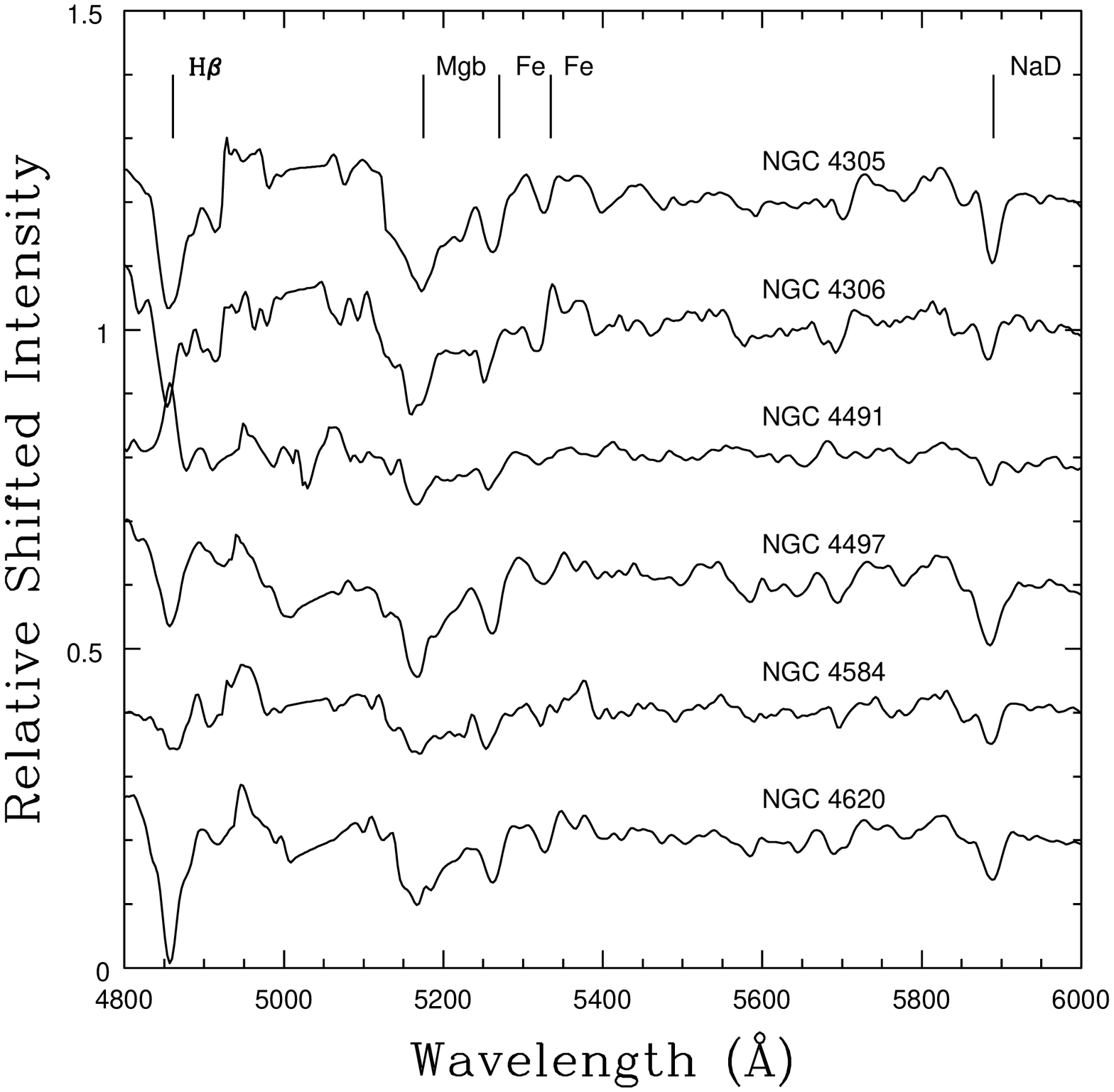}
\caption{GMOS-S spectra of the galaxy centers at visible wavelengths. 
The spectra have been normalized to the continuum 
and then shifted along the vertical axis for the purposes 
of display. Various atomic features are indicated. There is a galaxy-to-galaxy 
dispersion in the depths of many absorption features, as might be
expected given the range in central $J-K$ colors (Section 3). 
H$\beta$ is in emission in the NGC 4491 spectrum, and emission likely fills 
H$\beta$ in the NGC 4584 spectrum. The metallic absorption features in 
the NGC 4491 and NGC 4584 spectra are shallower than in the other galaxy spectra, 
likely due to luminous early-type stars and veiling by continuum emission.}
\end{figure*}

\begin{figure*}
\figurenum{4}
\epsscale{1.0}
\plotone{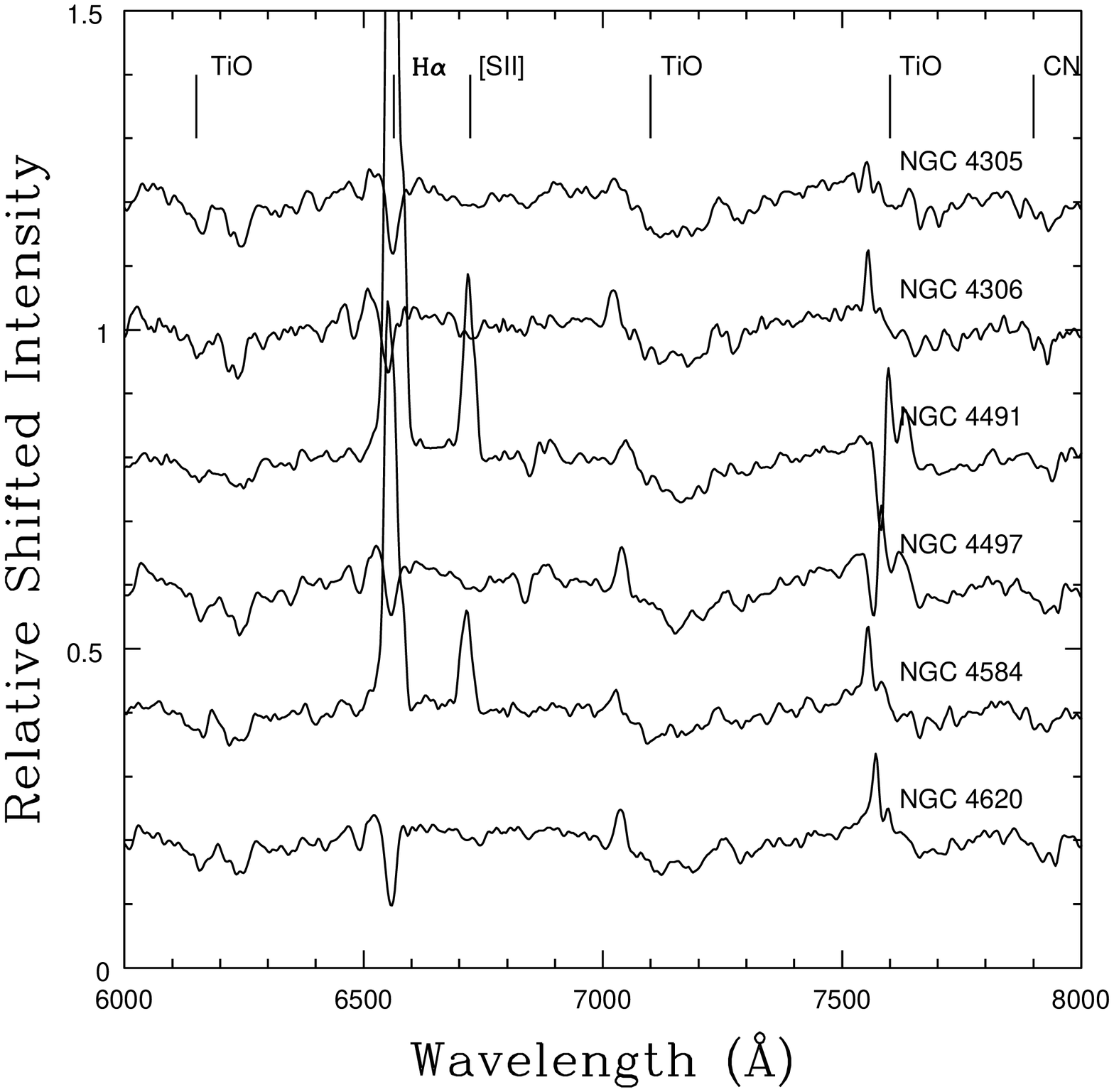}
\caption{Same as Figure 3, but covering wavelengths between 0.60 -- $0.80\mu$m. 
H$\alpha$ and [SII] emission lines are seen in the NGC 4491 and NGC 4584 spectra.}
\end{figure*}

\begin{figure*}
\figurenum{5}
\epsscale{1.0}
\plotone{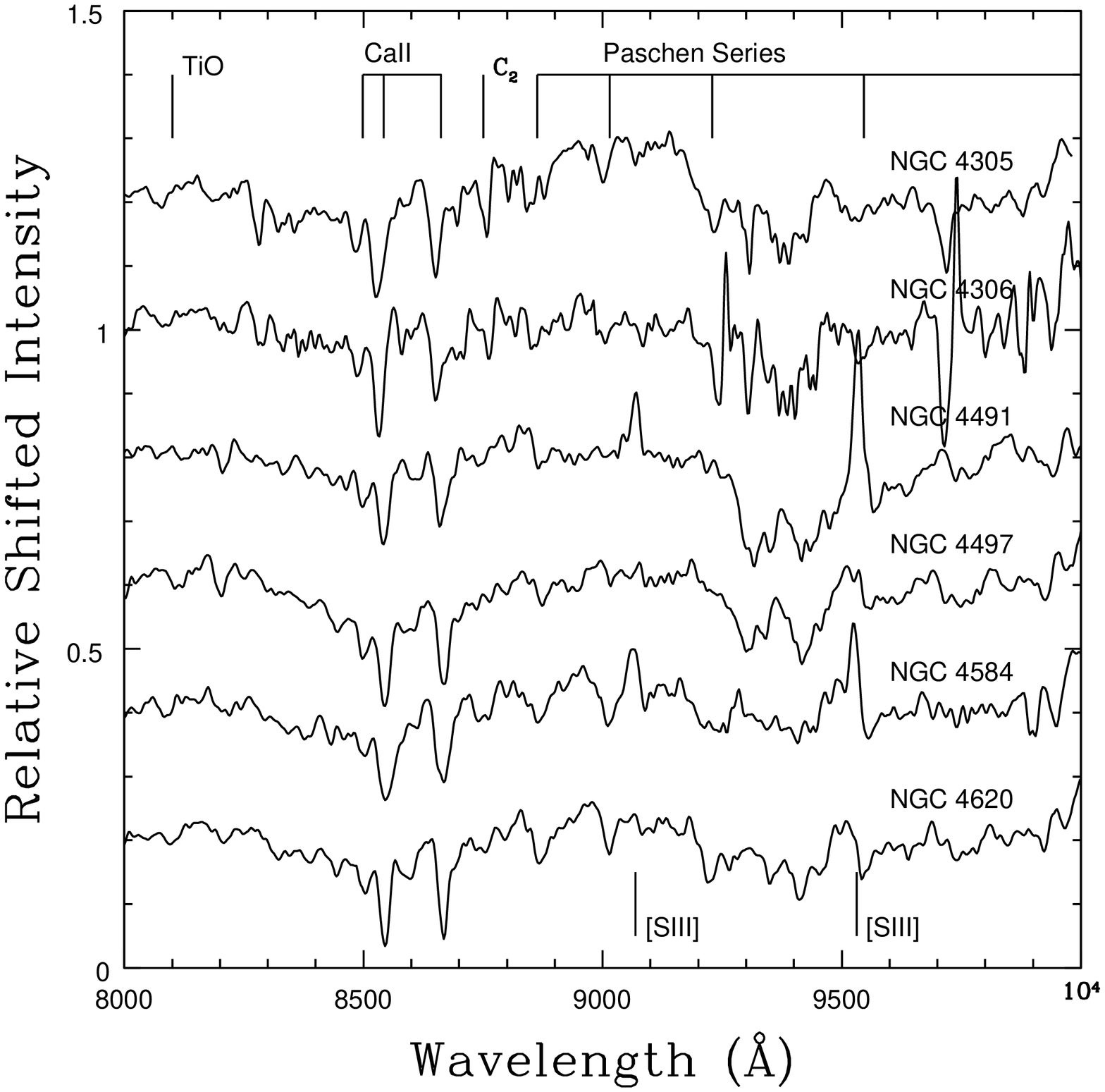}
\caption{Same as Figure 3, but covering wavelengths 0.80 -- $1.00\mu$m. The wavelengths 
of the reddest Paschen lines are indicated; the Paschen break occurs at 8204\AA . 
[SIII]$9069+9532$ emission lines are seen in the spectra of NGC 4491 and NGC 4584.}
\end{figure*}

	In the absence of emission, H$\alpha$ and H$\beta$ are basic 
probes of age. The Balmer lines are deepest in the spectrum 
of NGC 4620. However, the depths of these lines 
can be skewed by line emission, and H$\alpha$ and H$\beta$ are both in 
emission in NGC 4491. H$\alpha$ is also in emission in NGC 4584, and the weak nature of 
H$\beta$ absorption in the NGC 4584 spectrum suggests that emission fills that line. 
It is unlikely that low-level H$\alpha$ emission lurks in the spectra 
of the other galaxies, as ages that are estimated for these galaxies 
from H$\alpha$ agree with those found from H$\beta$ (Section 6).

	Emission from [SII]6746, [SIII]9069, and [SIII]9532 is 
seen in the NGC 4491 and NGC 4584 spectra, but 
neither [OIII] 4959 nor 5007 is detected. The ratio of [SII]6746 to H$\alpha$ 
provides insights into the source that powers 
this emission. Kewley et al. (2001) model emission from star-forming 
systems, and their Figure 6 indicates that when [OIII]/H$\alpha$ 
is small -- as is the case here -- then emission from systems in which 
log([SII]/H$\alpha) \leq -0.5$ is powered by radiation from hot stars, rather 
than from a LINER or an AGN. As log([SII]/H$\alpha) \leq -0.7$ in both NGC 4491 
and NGC 4584, then the emission from the central regions of these galaxies is 
likely powered by photons from luminous hot stars. GALEX images of both galaxies reveal 
bright cores in the FUV and NUV channels.

	Absorption lines that are familiar probes of 
metallicity and chemical mixture are present in Figures 3 -- 5. 
Metallic lines tend to be weakest in the spectra of NGC 4491 and NGC 
4584, likely due to the presence of luminous young stars and veiling by 
continuum emission. The deepest Fe line at 5270\AA\ is found in the spectra of 
NGC 4305 and NGC 4497. The lines of the Ca triplet are among the strongest 
features in the red spectra of early-type galaxies, and these 
trace metallicity in systems that have intermediate and old ages (e.g. Armandroff 
\& Zinn 1988). Deep Ca lines are evident in all the galaxy spectra in Figure 5. 

	The NaD doublet blends to form a single feature at the resolution of the 
GMOS-S spectra. NaD is deepest in NGC 4305 and NGC 4497, and is shallowest in NGC 4491.
NaD absorption may originate not only in stellar photospheres but also in interstellar 
material along the line of sight. While it is likely that NaD is dominated 
by the photospheric component in most of the spectra, an exception 
might be NGC 4491, the central regions of which may contain 
significant amounts of dust (Section 3). Assuming that the red central colors of 
NGC 4491 are due to dust and that the intrinsic $J-K$ of this galaxy is the 
same as in the other systems, then the color measurements in Figure 2 suggest that 
E(J--K) $\sim 0.2$ magnitudes. This corresponds to E(B--V) $\sim 0.4$ 
magnitudes if the Cardelli et al. (1989) extinction law is applied. Poznanski et al. 
(2012) calibrate the relation between the equivalent width of NaD and $E(B-V)$ 
using Galactic stars, and a color excess E(B--V) = 0.4 corresponds to a NaD 
equivalent width of $\sim 0.6$ \AA\ . This is roughly consistent with 
the equivalent width of NaD in the central spectrum of NGC 4491, hinting that at 
least some of the NaD absorption may be interstellar in origin. 

	Comparisons with model spectra are made in Section 6 and -- 
aside from the central regions of NGC 4491 and NGC 4584 -- these 
comparisons indicate that the bulk of the light from the centers of these galaxies 
originates from stars with ages $\sim 1.5 - 3$ Gyr. This is an age range where C 
stars may contribute significantly to the total light output (e.g. Maraston 2005). 
Indeed, Davidge (2018) examined the spectrum of the 2 Gyr old LMC 
cluster NGC 1978, and found that C stars produce a significant $0.79\mu$m CN 
feature in the integrated cluster spectrum. This feature is not an 
unambiguous C star indicator, as it is also present in the spectrum of evolved 
oxygen-rich stars, including RSGs (Davidge 2018). 
This CN band is detected in all six galaxies. 

\subsection{GMOS-N}

	The spectra recorded with GMOS-N have a higher resolution 
and extend to shorter wavelengths than the GMOS-S spectra, and so 
contain information that is complementary to the GMOS-S spectra.
The GN and GS spectra of the central regions of NGC 4305 and NGC 4620 
are compared in Figures 6 and 7. Absorption lines in 
the GN spectra are deeper and sharper than in their GS counterparts, 
reflecting the higher wavelength resolution of the GMOS-N spectra. 

	There are obvious similarities between the two sets of spectra 
in Figures 6 and 7. This similarity is of particular importance for 
NGC 4305, as the GMOS-N spectra of this galaxy were acquired by 
stepping the galaxy between two locations on the slit, allowing for better suppression 
of telluric emission features than was the case for the GMOS-S spectra of 
this same galaxy. The comparisons in Figure 7 validate the GMOS-S spectrum of NGC 4305. 
Residuals from telluric emission lines do not skew line strengths 
in the GS spectrum of NGC 4305, and this is confirmed in the next section where 
the line indices obtained from the GMOS-N and GMOS-S spectra are compared. 

\begin{figure*}
\figurenum{6}
\epsscale{1.0}
\plotone{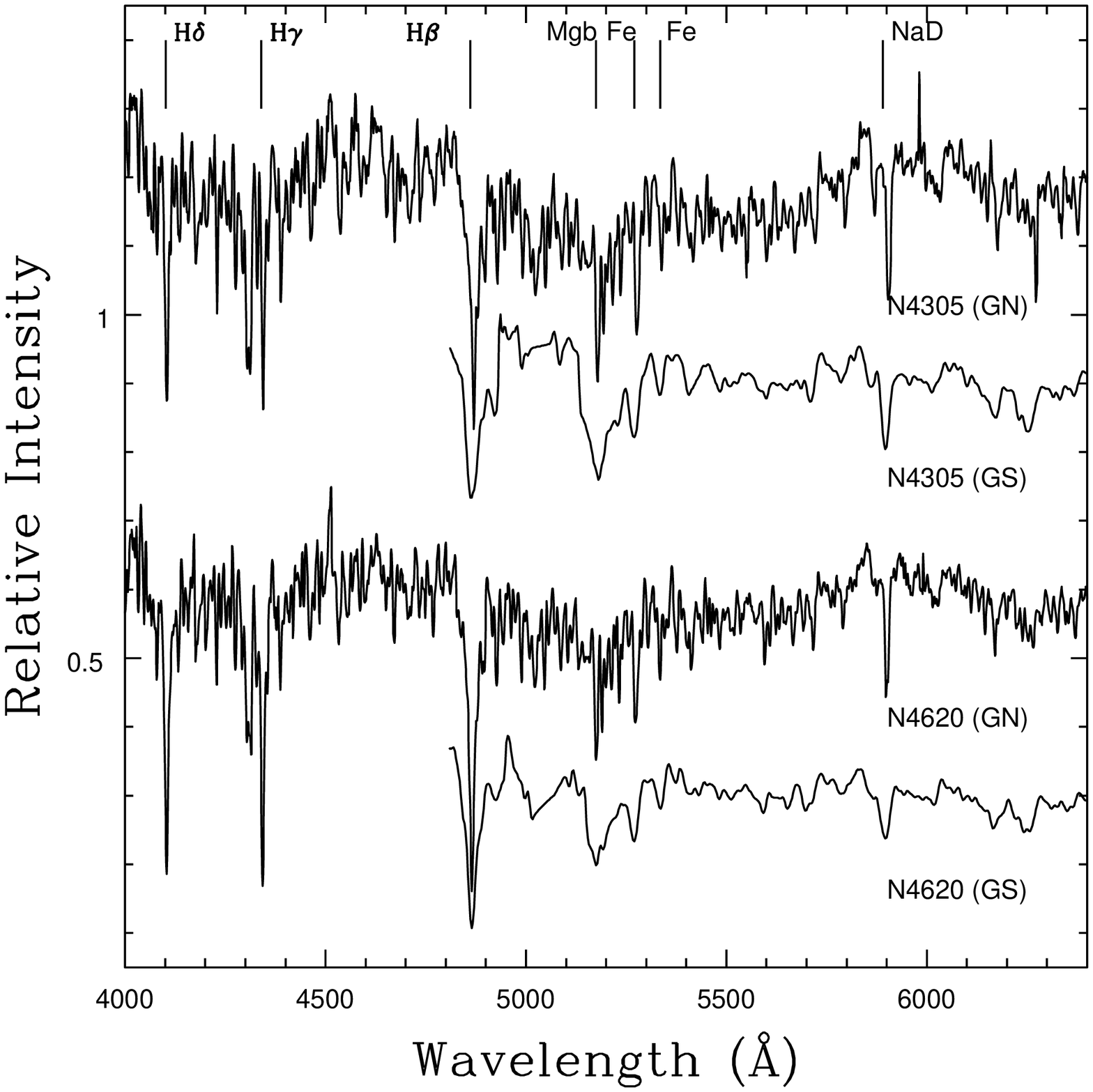}
\caption{GN and GS spectra of the central regions of NGC 4305 
and NGC 4620 in the wavelength interval observed with 
the B600 grating. The different wavelength resolutions of the 
GS and GN spectra are obvious, and the shorter wavelength coverage of the 
GN spectra allows H$\gamma$ and H$\delta$ to be detected. The higher 
spectral resolution of the GN data also should make emission lines that are 
hidden in absorption features easier to detect. There is no evidence of H$\beta$ 
emission in the GN spectra.}
\end{figure*}

\begin{figure*}
\figurenum{7}
\epsscale{1.0}
\plotone{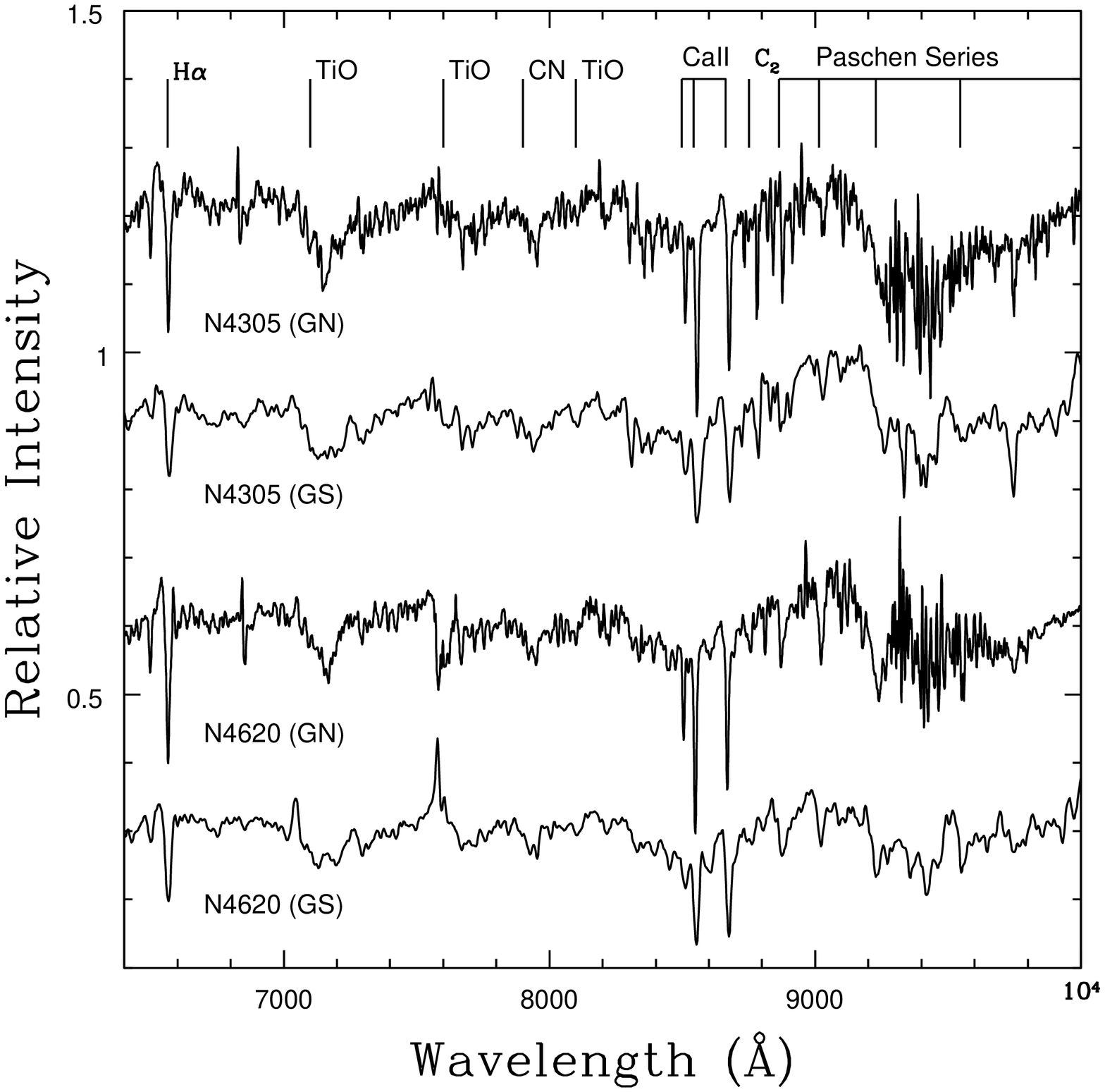}
\caption{Same as Figure 6, but covering the wavelength interval 
observed with the R400 grating. The noise between 
9200 and 9500\AA\ in the GN spectra is due to reduced atmospheric transmissivity 
caused by telluric H$_2$O. That the GN and GS spectra agree at these wavelengths 
suggests that the large-scale structure of the telluric H$_2$O feature has been 
successfully removed. The depression in the spectra between 0.92 and $1.00\mu$m is 
intrinsic to the galaxies, and is due to TiO and/or CN. There is no 
evidence for H$\alpha$ emission in the GN spectra of either galaxy.}
\end{figure*}

	The higher resolution of the GN spectra should facilitate the detection of low 
level emission in the Balmer lines, and no evidence for emission is seen in H$\alpha$ 
and H$\beta$ in the GN spectra. The higher resolution of the GN spectra also 
makes it easier to identify weak features, or those that are 
blended with other lines at lower resolution. However, line identifications can 
be ambiguous. For example, there is a C$_2$ band head near 8750\AA\ , and a 
feature near this wavelength is present in the GMOS-N spectra of NGC 4305 and NGC 4620. 
However, a Paschen series line also occurs at this wavelength. 
Given the prominent Balmer lines in the spectra of these galaxies, it is 
likely that the feature in question is due to hydrogen, and not C$_2$.

\section{LINE INDICES AND GRADIENTS}

	Spectral indices are one means of quantifying the depths of 
absorption and emission features in galaxy spectra. 
The majority of the indices used in the present study 
are in the Lick system, as summarized by Worthey et al. (1994). The Lick 
indices sample prominent features at visible wavelengths that 
are sensitive to metallicity, chemical mixture, and age.
As noted by Worthey et al. (1994), these indices typically do not 
measure signal from a single element or molecule. While the targeted wavelength range 
contains a dominant atomic or molecular feature by definition, there are 
inevitably transitions from other species present. Likewise, wavelength 
intervals defined to measure the continuum are also subject to contamination. 
The indices discussed here are in an instrumental system, and most were selected for 
study because they have been shown to have a low sensitivity to spectral resolution. 
Indices in the same instrumental system are measured from model spectra, and 
compared with the observations in Section 6. 

	Puzia et al. (2013) examine the transformation of indices measured from 
GMOS spectra using the B600 grating into the Lick system. The B600 grating 
was used for the GMOS-N spectra of NGC 4305 and NGC 4620, thereby 
allowing a body of internally consistent indices to be selected 
from those examined by Puzia et al. (2013) that enable 
reliable galaxy-to-galaxy and intergalaxy (i.e. line gradient) comparisons. 
Indices measured for NGC 4305 and NGC 4620 in this internal system can then be 
used to calibrate measurements in the GMOS-S spectra (Section 5.2).

	The transformation coefficients listed in Tables 
4 ($2 \times 2$ pixel binning; 0.75 arcsec slit) and 5 ($1 \times 1$ 
binning; 0.5 arcsec slit) of Puzia et al. (2013) can be used to 
identify indices that have a low sensitivity to spectral resolution. 
The GMOS-N spectra of NGC 4305 and NGC 4620 were recorded 
with a 1 arcsec slit, and thus have a lower resolution than those used to construct 
Table 4 of Puzia et al. (2013). This difference notwithstanding, the resolution of the 
NGC 4305 and NGC 4620 spectra is closer to that of the native Lick system than the 
configuration used to produce the entries in Table 4 of Puzia et al. (2013). 

	A comparison of the entries in Tables 4 and 5 of Puzia et al. (2013) 
allows indices to be identified that (1) have transformation 
coefficients that agree at the $2\sigma$ level, and (2) monitor features that 
provide metallicity and age information. The indices that are 
measured in the GMOS-N spectra are: CN$_2$, G4300, H$\beta$, Mg$_2$, 
Fe5270, Fe5335, NaD, and TiO$_2$. A sub-set of these are measured in the GS 
spectra (Section 5.2). Three of the indices sample broad 
molecular features, and thus have a wide wavelength coverage that 
makes them largely immune to changes in spectral resolution. The other indices 
measure some of the strongest atomic features in galaxy spectra at visible wavelengths. 

	Other indices were also measured. The CaT index defined by Cenarro et 
al. (2001) probes metallicity in intermediate age and old populations. 
Smoothing experiments discussed by Cenarro et al. (2001) and Davidge (2018) show 
that the CaT index varies slowly with respect to changes in spectral resolution. 
The Ca triplet has been found to be most sensitive as 
a metallicity indicator among systems with [M/H] $< -0.5$, 
although the metallicity sensitivity continues to higher metallicities among SSPs with 
ages $< 5$ Gyr (Figure 5 of Vazdekis et al. 2003). 
Still, the utility of the Ca triplet as a metallicity tracer in complex stellar 
systems may be subject to uncertainties in the chemical enrichment history of Ca. 
There is only a modest galaxy-to-galaxy dispersion in the depths of the Ca triplet 
lines when systems with a range of masses are compared. This could 
reflect a fixed [Ca/H] among the galaxies (e.g. Thomas 
et al. 2003a), and/or saturation in the metallicity sensitivity of Ca triplet lines 
among old, metal-rich populations. 

	Jones \& Worthey (1995) and Worthey and Ottaviani (1997) discuss indices that 
measure the depths of H$\gamma$ and H$\delta$. These provide an independent measure of 
age, and so can be used to assess the influence of line emission on ages estimated 
from lower order Balmer lines. The entries in Tables 4 and 5 of Puzia et al. (2013) 
indicate that H$\gamma_{A}$ is a relatively robust index in terms of spectral 
resolution, and so it is measured in the GMOS-N spectra of NGC 4305 and NGC 4620. 

	Two indices that are not in a standard system but that monitor the 
strengths of astrophysically interesting features are also discussed. One of these 
features is H$\alpha$, which has the potential to constrain age, with the caveat that 
H$\alpha$ is more susceptible to contamination from line emission than higher order 
Balmer lines. In this study, the depth of H$\alpha$ is 
measured in the wavelength interval 6540 to 6590\AA\ . 
The continuum is measured in the intervals 6510 to 6540\AA\ and 6590 to 6620\AA\ . 

	Another index measures the depth of the CN band near 7900\AA\ , which 
Davidge (2018) found is sensitive to the presence of C stars in the integrated 
light of intermediate age simple stellar systems. This CN band has been used in 
the past to distinguish between M giants and C stars in narrow-band photometric surveys 
(e.g. Richer, Crabtree, \& Pritchet 1984; Cook, Aaronson, \& Norris 1986). 
The CN7900 index is measured with continua in the 
wavelength intervals 7800 -- 7875\AA\ and 8005 -- 8070\AA , 
while the depth of the CN band is measured between 7875 and 7975\AA . 
As with the other molecular indices discussed in this study, CN7900 measurements 
are in magnitude units. 

\subsection{Indices measured from the GMOS-N spectra of NGC 4305 and NGC 4620}

	The radial behaviour of the indices in NGC 4305 and NGC 4620 measured 
from the GMOS-N spectra is examined in Figures 8 -- 10. Indices were measured 
from mean spectra of each galaxy in the 5 regions listed in Table 3.
Figure 8 shows indices that are sensitive to age, 
while Figures 9 and 10 show indices that monitor metallicity and chemical mixture. 
Molecular indices are shown in Figure 9, and atomic indices in Figure 10. 
The Fe5270 and Fe5335 indices are summed in Figure 10 to increase the S/N ratio. 
Thomas et al. (2003b) define the hybrid [MgFe]' index as a measure of 
total metallicity, and this is plotted for each galaxy in the lower panel of Figure 10. 
The error bars show random errors deduced from the scatter in the index measurements, 
and do not include systematic effects. 

\begin{figure*}
\figurenum{8}
\epsscale{1.0}
\plotone{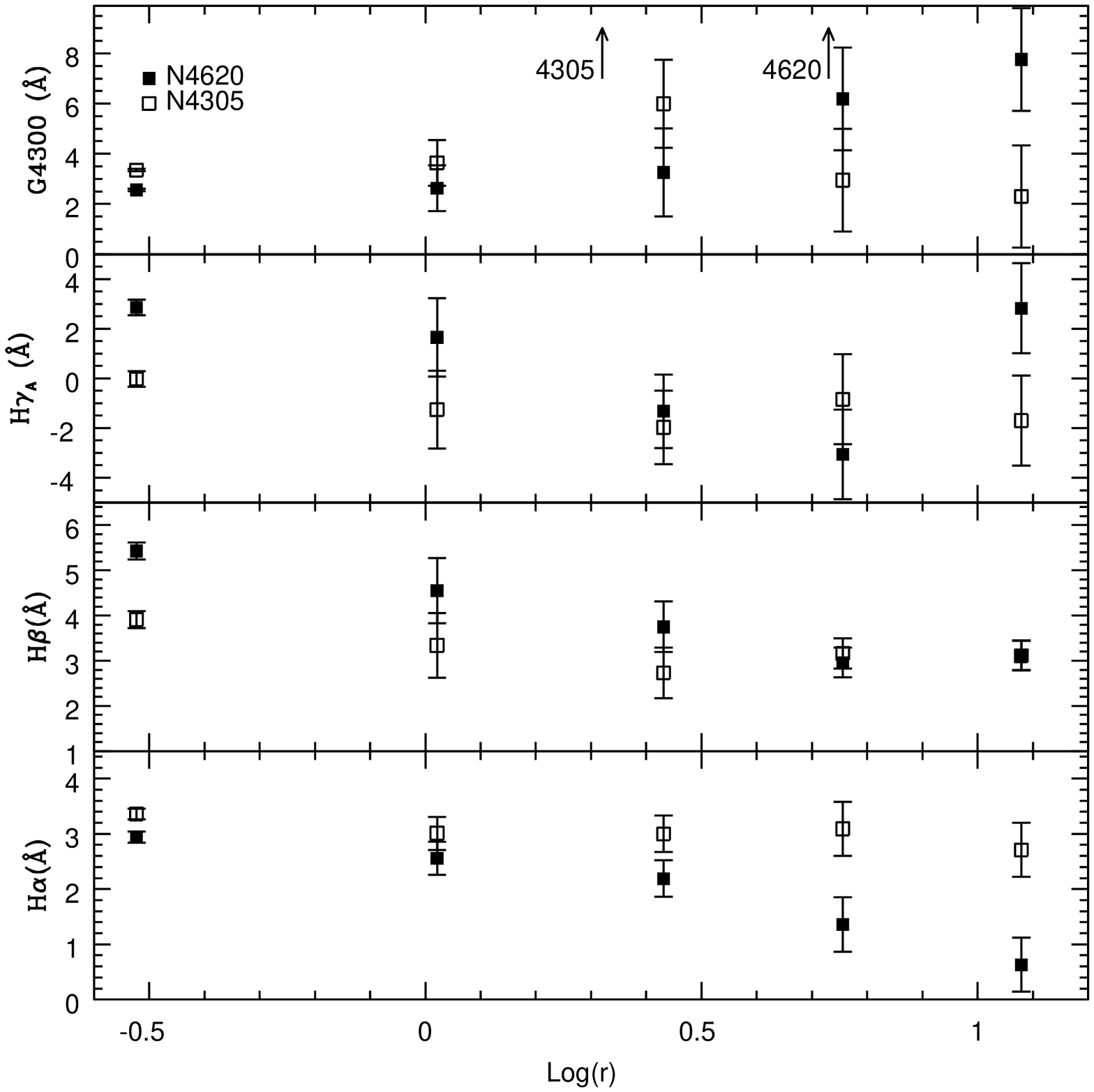}
\caption{Age sensitive indices measured from the GMOS-N spectra of NGC 4305 
(open squares) and NGC 4620 (filled squares). As in all subsequent figures (1) $r$ 
is the angular offset from the nucleus in arcsec along the slit, and (2) 
the half light radius of each bulge, as measured by McDonald et al. (2011) and 
corrected for slit orientation, is marked in the top panel. 
Note that the response of the G4300 and Balmer indices to changes in age 
differ: whereas the Balmer indices weaken with increasing age, 
G4300 strengthens as age increases. All four indices 
stay more-or-less constant with radius in NGC 4305. 
G4300 increases in strength at large r in NGC 4620, while H$\beta$ and H$\alpha$ 
decrease in strength.}
\end{figure*}

\begin{figure*}
\figurenum{9}
\epsscale{1.0}
\plotone{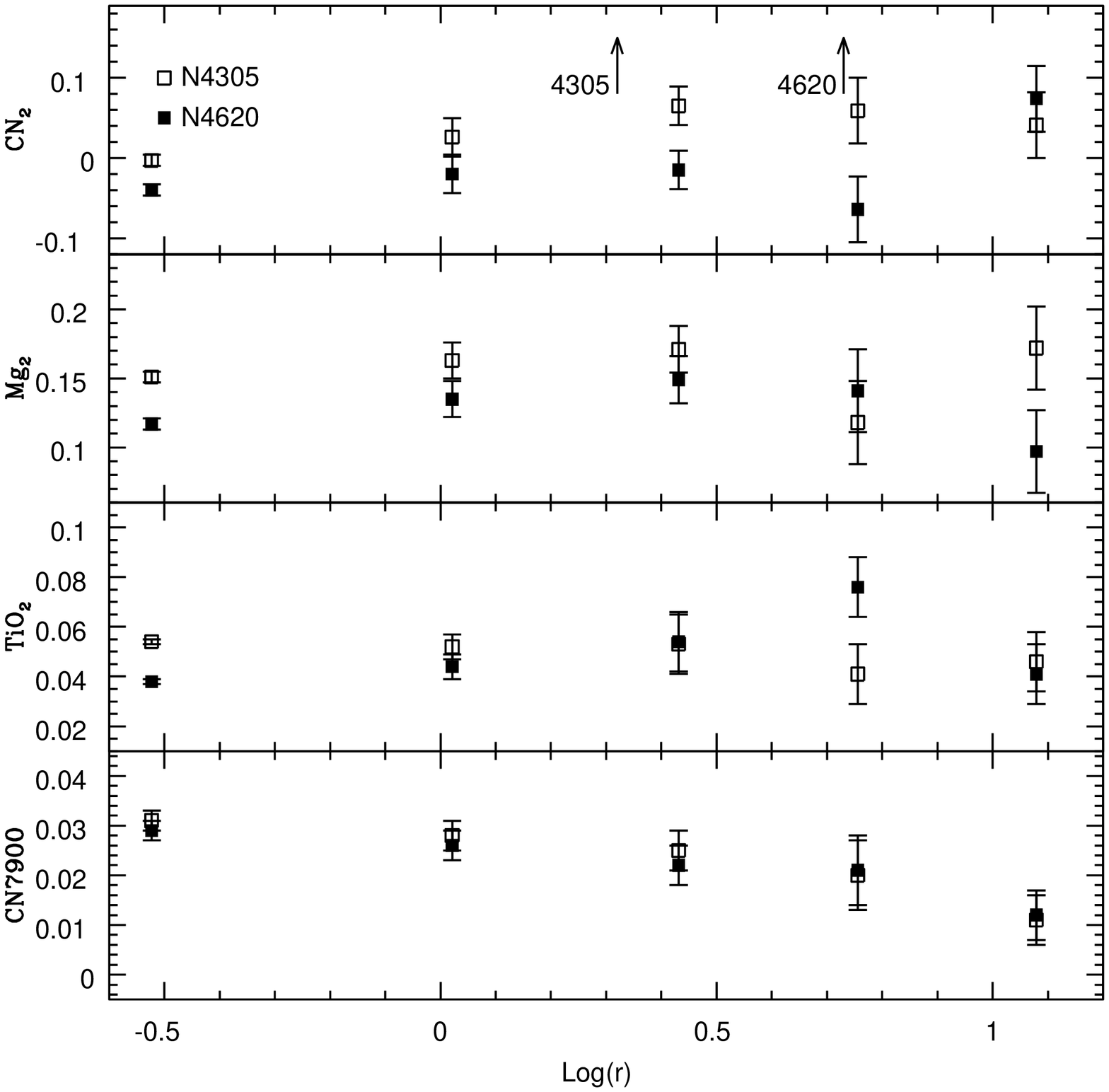}
\caption{Molecular indices measured from the GMOS-N spectra 
of NGC 4305 and NGC 4620. With the exception of CN7900, there 
is no evidence for systematic radial gradients in the molecular indices. 
The CN7900 indices of both galaxies are remarkably similar at all 
radii. The radial behaviour of Mg$_2$ in both 
galaxies is consistent with the behaviour of Mgb, shown in Figure 10.}  
\end{figure*}

\begin{figure*}
\figurenum{10}
\epsscale{1.0}
\plotone{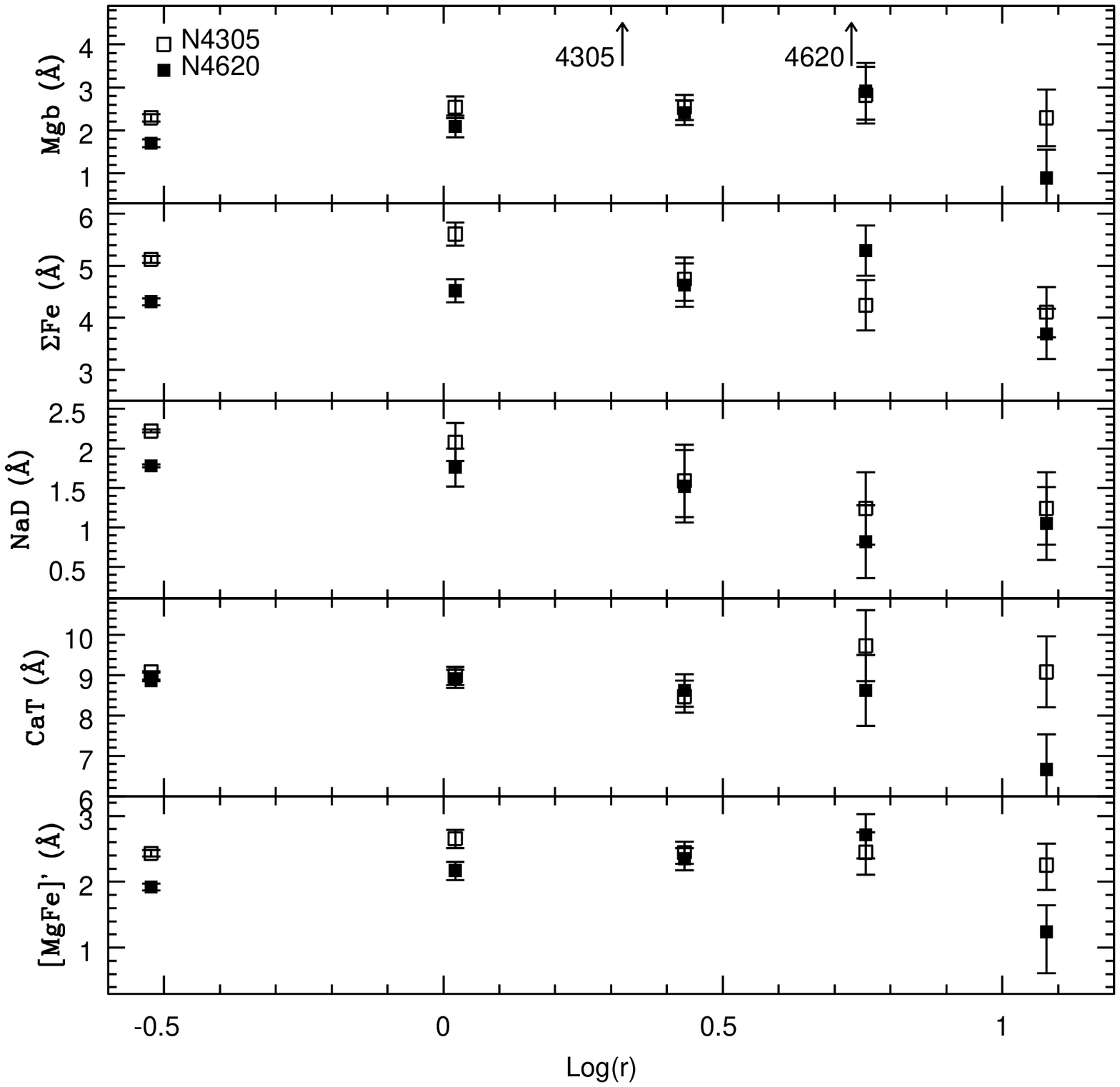}
\caption{Metal line indices measured from the GMOS-N 
spectra of NGC 4305 and NGC 4620. Fe5270 and Fe5335 
have been summed to increase the S/N ratio, and the result is $\Sigma$Fe. Mgb is 
more-or-less constant with radius in NGC 4305, and this is 
consistent with the behaviour of Mg$_2$ in Figure 9. There is a 
tendency for $\Sigma$Fe to weaken towards larger radii in NGC 4305, 
while $\Sigma$Fe is constant with radius in NGC 4620. 
Mgb and CaT in both galaxies show similar behaviour, including a large 
drop between log(r) = 0.75 and 1.1 in NGC 4620 -- this interval spans the 
transition from bulge to disk-dominated light in this galaxy. 
NaD weakens with radius in both galaxies. The hybrid [MgFe]' index 
(Thomas et al. 2003b) does not change in a systematic way with radius in NGC 4305, 
but drops at large radii in NGC 4620.} 
\end{figure*}

\subsubsection{Age Sensitive Indices}

	The radial behaviours of H$\gamma_A$, G4300, H$\beta$, and H$\alpha$ 
are examined in Figure 8. The G4300 and Balmer indices 
monitor (at least in part) the spectral type of stars near the main 
sequence turn-off (MSTO), and so provide age information. If metallicity 
remains fixed then as one moves to progressively 
younger ages the temperature of stars near the MSTO increases, and the 
spectral type moves to earlier values. This results in a strengthening 
of the Balmer lines, and a decrease in the depth of the G-band. 
The sensitivity of G4300 to age is thus in the opposite sense 
to that of the Balmer lines: whereas the Balmer lines weaken as age increases, 
G4300 strengthens with increasing age (e.g. Figure 42 of Worthey 1994).
G4300 and H$\gamma_A$ are of special interest as they are less susceptible to 
contamination from emission than H$\beta$ and H$\alpha$. 

	Considering the uncertainties in the indices, G4300 is 
more-or-less constant with radius in NGC 4305, and this is 
consistent with the behaviour of the three Balmer indices. 
As for NGC 4620, G4300 strengthens with increasing radius while H$\beta$ and H$\alpha$ 
weaken with increasing distance from the galaxy center. H$\gamma_A$ also 
weakens with increasing radius when log(r) $\leq 0.75$, which is the 
part of the galaxy that is dominated by bulge light, although it jumps 
to higher values in the last bin. To the extent that these indices track age 
then -- with the exception of the H$\gamma_A$ measurement in the outer most 
regions of NGC 4620 -- the central regions of NGC 4620 may be 
younger than the surroundings.

\subsubsection{Indices that Probe Metallicity}

	In Section 3 it was shown that $J-K$ decreases 
with increasing radius in NGC 4305, but not in NGC 4620. 
If $J-K$ tracks metallicity in these galaxies then the metallicity indices in NGC 
4305 and NGC 4620 might show different radial behaviours.
Molecular indices measured from the NGC 4305 and NGC 4620 GMOS-N spectra are 
shown in Figure 9. There is a tendency for CN$_2$, Mg$_2$, and TiO$_2$ to 
remain more-or-less constant with radius in both galaxies, although Mg$_2$ drops at 
large radii in NGC 4620. CN7900 is the only molecular 
index in Figure 9 that shows a systematic gradient 
in both galaxies, and there is remarkable agreement 
between the CN7900 measurements in both galaxies at all radii. This agreement is 
consistent with the modest dispersion in CN7900 obtained from the GS spectra of 
galaxies that do not show line emission, and this is discussed further in 
Section 5.2.

	Indices that examine the strengths of atomic metal lines are shown 
in Figure 10. The Mgb and CaT indices define similar radial trends in each galaxy. 
There is not a systematic radial gradient in the Mgb and CaT indices in NGC 4305, 
and this is consistent with the behaviour of the Mg$_2$ index. The 
Mgb and CaT indices drop at large radii in NGC 4620, and a similar dip is seen in 
the Mg$_2$ index of this galaxy in Figure 9.

	There is a tendency for $\Sigma$Fe to weaken towards larger radii in NGC 
4305, although the significance of this result is not clear given the error bars. 
As for NGC 4620, $\Sigma$Fe stays more-or-less constant or may even rise 
with increasing radius, before dropping at larger radii. Similar behaviour 
is seen in the NGC 4620 Mgb measurements. 
The hybrid [MgFe]' index stays constant throughout NGC 4305, while [MgFe]' appears 
to climb with increasing radius in NGC 4620 before dropping in the outermost 
regions. This parallels the radial behaviour of the Mgb and $\Sigma$Fe indices.

	NaD was found to show steep, well-defined, gradients in the 
spectra of the early-type galaxies examined by Davidge (1992). NaD noticeably 
weakens with increasing radius in both NGC 4305 and NGC 4620. Given that other 
indices do not change with radius in NGC 4305, and may even climb with 
increasing radius in NGC 4620, then either the Na abundance in 
stellar photospheres does not track the abundances of other elements, 
and/or the depth of NaD may be affected by interstellar absorption. 
As the $J-K$ colors of these galaxies are not peculiar then 
the first explanation is favored. La Barbera et al. (2017) investigate the 
behaviour of multiple Na lines over a wide wavelength interval in the spectra of 
two moderately massive early-type galaxies. They find that while [Na/Fe] is 
more-or-less constant near the center of one galaxy, it drops at larger radii. 
They also find that NaD is more sensitive to [Na/Fe] than other Na I features at 
red and NIR wavelengths. Hence, [Na/Fe] may change 
with radius in NGC 4305 and NGC 4620.

	In summary, with the exception of NaD and CN7900, there is no evidence for 
gradients in metallic features in NGC 4305. As for NGC 4620, metallic indices 
appear to stay constant or possibly even rise with increasing radius, before 
dropping at log(r) $> 0.75$. The age-sensitive indices in the spectra of 
NGC 4305 reveal no hint of a gradient. However, there is 
evidence for a gradient in NGC 4620, in the sense that the center has deeper 
Balmer lines than the surroundings. An age gradient may complicate 
the interpretation of the metallic features in NGC 4620, and might even 
explain the negative atomic line gradients that are present when 
log(r) $\leq 0.75$ in this galaxy.

\subsection{Indices measured from the GMOS-S Spectra}

	The GMOS-S spectra have a lower wavelength resolution than the GMOS-N spectra, 
and this compromises the ability to measure the depths of even moderately strong atomic 
features. Here, emphasis is placed on measuring 
indices from the GMOS-S spectra that have a wide wavelength coverage and/or that 
sample deep, isolated features. Metallicity in the 
GMOS-S spectra is then probed using the Mg$_2$, TiO$_2$, CN7900, and CaT indices. The 
NaD index is also measured as this traces a deep, isolated feature that 
exhibits well-defined gradients in other galaxies (Davidge 1992). Finally, 
the H$\beta$ and H$\alpha$ indices are also examined to allow age to be examined.

	The indices measured from the GMOS-S spectra are placed into the 
instrumental system defined by the GMOS-N spectra by comparing the indices obtained for 
NGC 4305 and NGC 4620 from both telescopes, computing an average offset for each 
index, and then applying the result to the GS index. While this is the simplest 
transformation possible, it is justified given that the spectroscopic properties 
of NGC 4305 and NGC 4620 are similar to those of NGC 4306 and NGC 4497 at all radii, 
as well as in the outer regions of NGC 4491 and NGC 4584. 
The Mg$_2$, TiO$_2$, and CN7900 indices for NGC 4305 and NGC 4620 measured from 
GMOS-N and GMOS-S agree to within a few thousandths of a magnitude. This excellent 
agreement is not surprising given the expected low sensitivity to spectral 
resolution for indices that span a broad wavelength range. As for the NaD and 
H$\alpha$ indices, the differences between the GMOS-N and GMOS-S measurements 
is 0.3\AA , in the sense that the GMOS-N indices are larger. The difference between 
the GN and GS CaT and H$\beta$ indices are 2.0\AA\ (CaT) and 1.4\AA\ (H$\beta$).

	The radial behaviour of indices measured in the GMOS-S 
spectra are examined in Figures 11, 12, and 13. Figures 11 and 12 show 
indices for the galaxies that do not have line emission, while 
Figure 13 shows indices for NGC 4491 and NGC 4584. 
Indices measured from the GMOS-N spectra of NGC 4305 and 
NGC 4620 are also plotted in Figures 11 and 12, and there is good agreement 
with the GMOS-S indices.

\begin{figure*}
\figurenum{11}
\epsscale{1.0}
\plotone{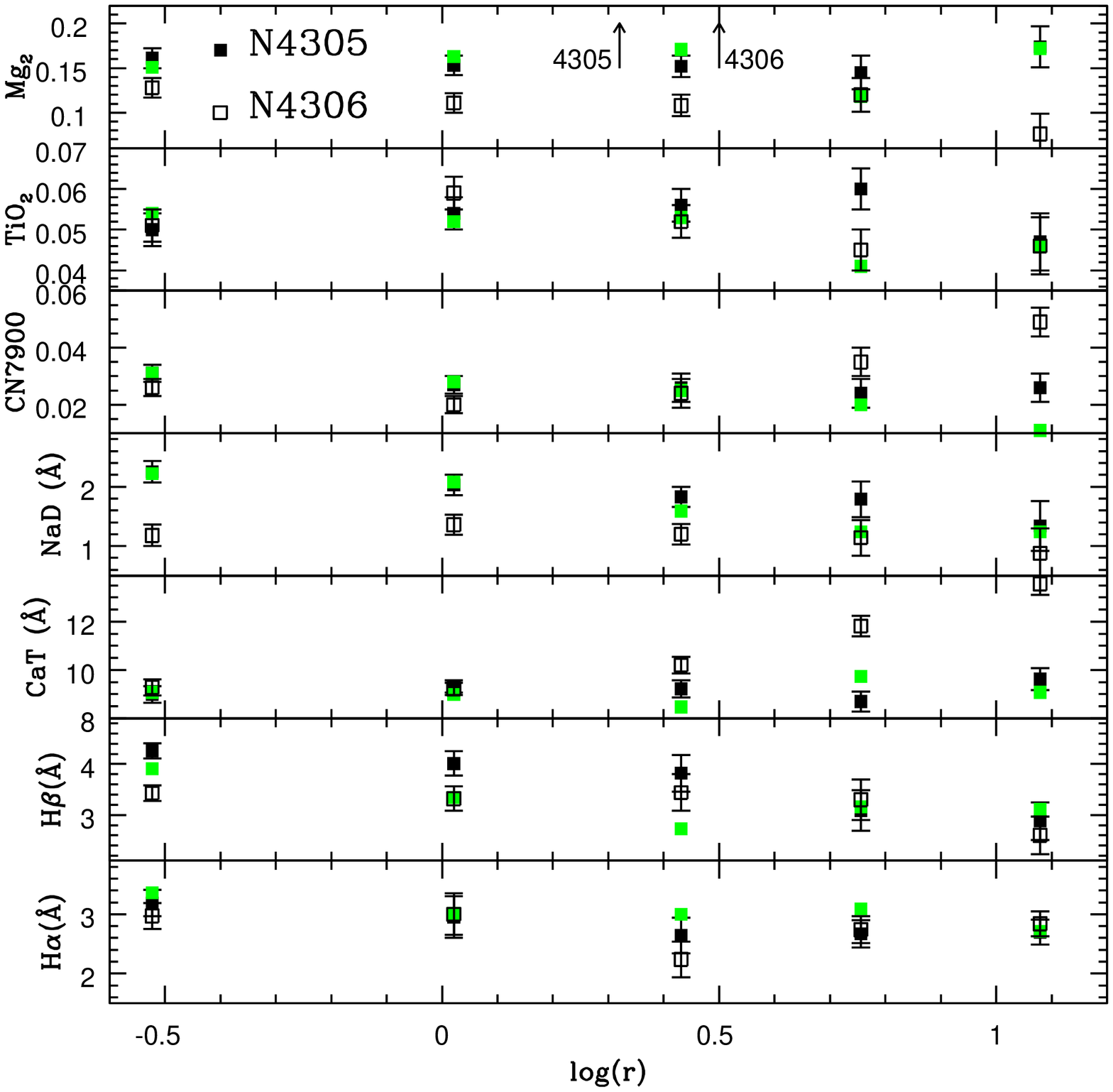}
\caption{Indices measured from the GS spectra of NGC 4305 and NGC 4306. 
The green squares are indices from the GMOS-N spectrum of NGC 4305, and 
there is general agreement between the GN and GS indices. 
The Mg$_2$ and TiO$_2$ indices in both galaxies are more-or-less 
constant with radius, while there is a tendency for NaD to weaken 
with increasing radius in both galaxies. CaT stays constant with radius in NGC 
4305. The tendency for CaT to strengthen in the outer regions of 
NGC 4306 is due to the imperfect subtraction of telluric 
emission lines. There is a tendency for H$\beta$ to weaken with increasing radius 
in both galaxies, although this is not the case for H$\alpha$. 
Comparisons with models point to significant age and metallicity gradients 
in NGC 4305 (Section 6).} 
\end{figure*}

\begin{figure*}
\figurenum{12}
\epsscale{1.0}
\plotone{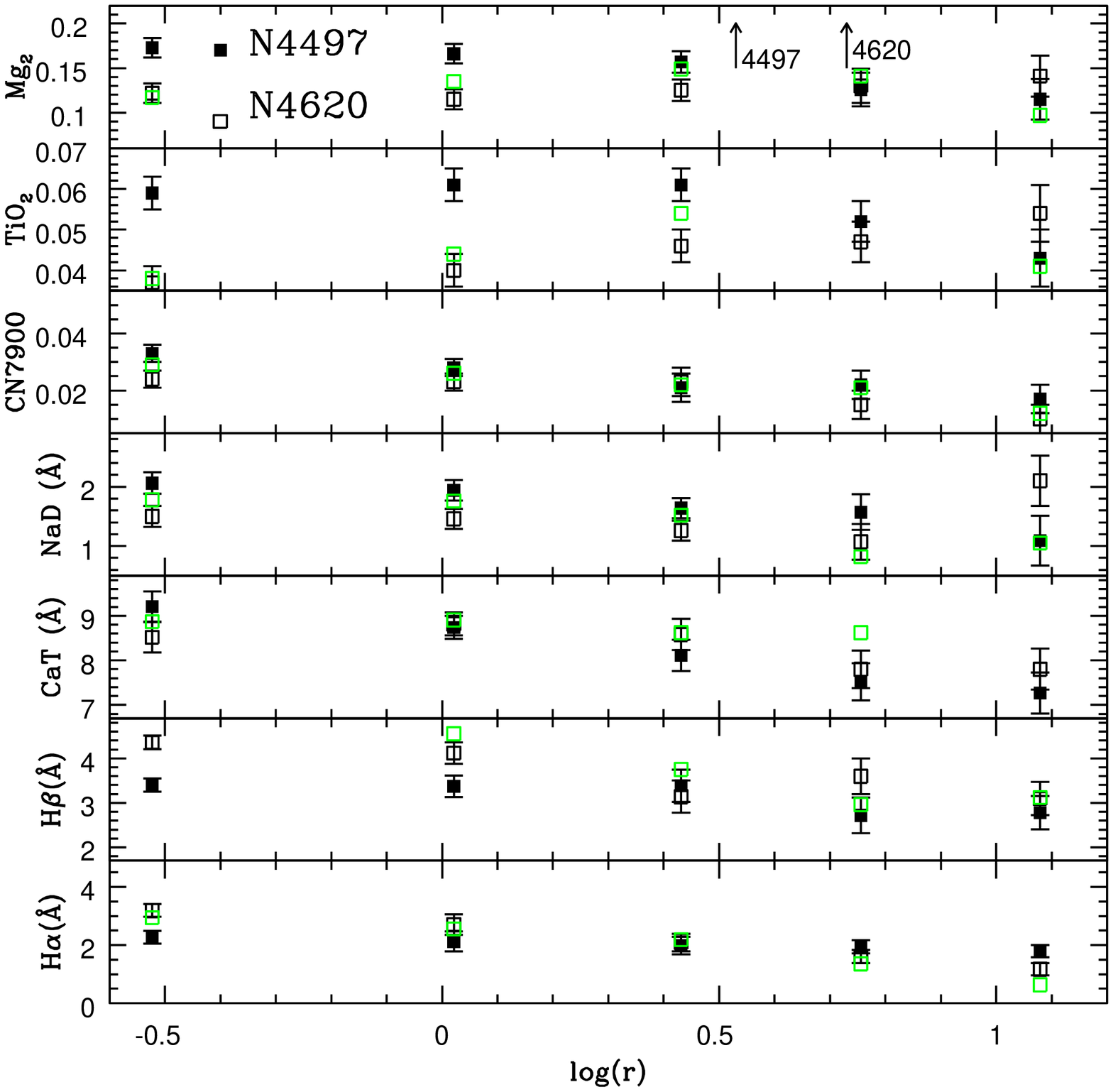}
\caption{Same as Figure 11, but showing indices for 
NGC 4497 and NGC 4620. The open green squares are 
indices measured from the GMOS-N spectra of NGC 4620. As in Figure 11, 
there is good agreement between the GN and GS indices. 
There is a tendency for both H$\beta$ and H$\alpha$ to weaken with increasing 
radius in NGC 4497 and NGC 4620. The Mg$_2$, TiO$_2$, NaD, and CaT indices weaken with 
increasing radius in NGC 4497. However, these same indices behave differently 
in NGC 4620: Mg$_2$ stays more-or-less constant, while there 
may be a negative TiO$_2$ gradient in the GMOS-S measurements. 
The CaT index weakens with increasing radius in NGC 4620.}
\end{figure*}

\begin{figure*}
\figurenum{13}
\epsscale{1.0}
\plotone{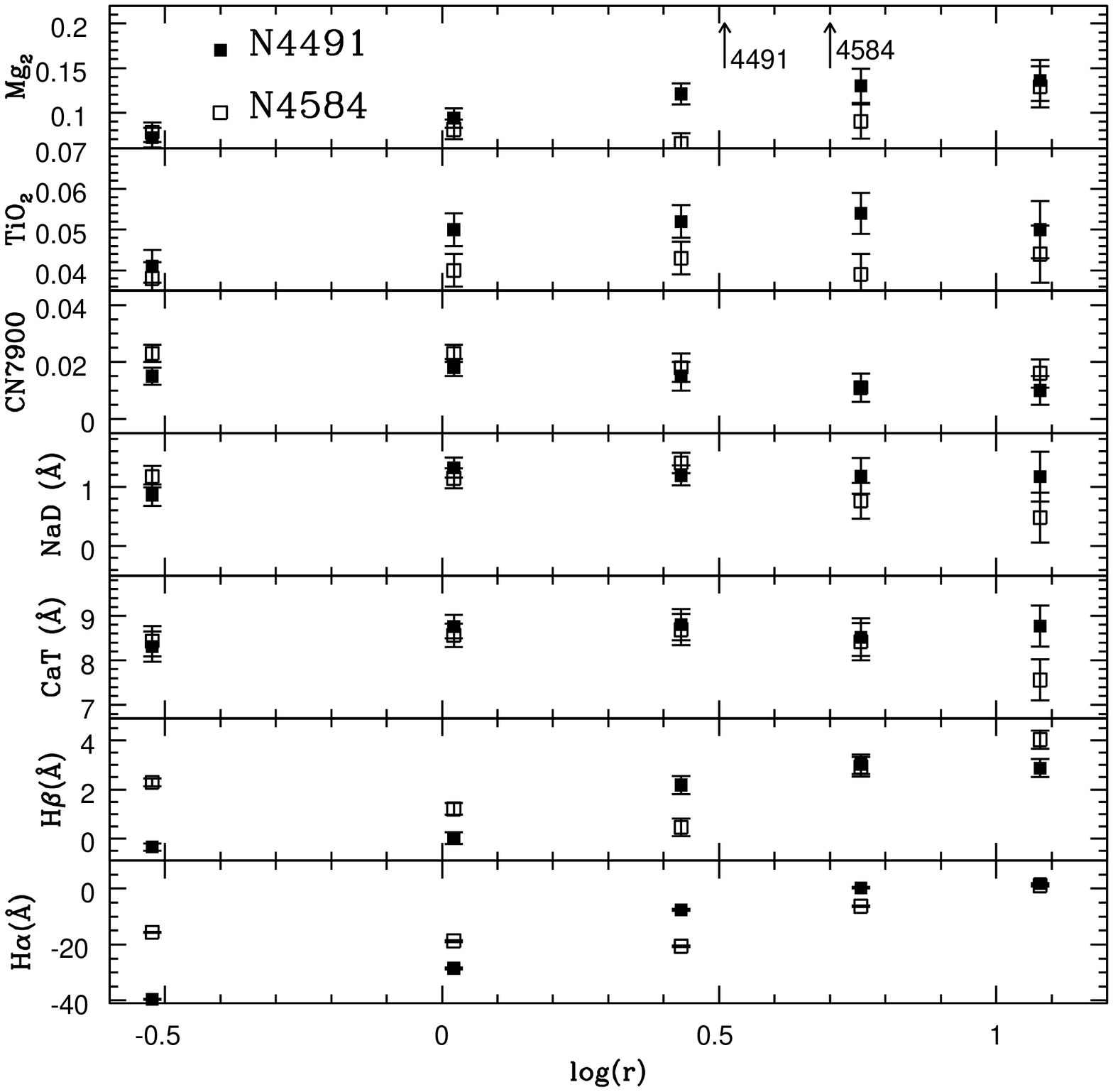}
\caption{Same as Figure 11, but showing indices measured in the GS spectra of 
NGC 4491 and NGC 4584. Many of the indices that probe metallic features in these 
galaxies have negative gradients or stay constant with radius, due to 
radial changes in the relative contribution made to the light 
by the largely featureless continuum from young stars and nebulosity. 
The CaT indices of both galaxies are similar over a large range of radii. The CaT and 
CN7900 features at small radii likely originate in RSGs. The H$\beta$ indices in 
the last two radial bins of NGC 4491 and NGC 4584 are comparable to those in NGC 4305, 
NGC 4497, and NGC 4620, raising the possibility that star formation in the disks of 
NGC 4491 and NGC 4584 may have been quenched at similar epochs to when 
star formation was curtailed in the other galaxies.}
\end{figure*}

\subsubsection{Mg$_2$, TiO$_2$, and CaT}

	Mg$_2$ is more-or-less constant with radius in the GMOS-S spectra of NGC 
4305 and NGC 4620, in agreement with the measurements made from the GMOS-N spectra. 
There is a pronounced Mg$_2$ gradient in NGC 4497, which also has the largest central 
Mg$_2$ index of the six galaxies. NGC 4491 and NGC 4584 
have the weakest central Mg$_2$ indices, due to veiling by the 
blue continuum associated with the young stars that power the line emission at 
small radii in these galaxies. The negative Mg$_2$ gradients in NGC 4491 and NGC 4584 
reflect the changing radial contributions made by continuum emission.

	The radial trends in the TiO$_2$ 
indices more-or-less parallel those in Mg$_2$, albeit with 
smaller radial variations than defined by Mg$_2$. 
NGC 4620 appears to be an exception, as the GMOS-S TiO$_2$ index in that 
galaxy has a negative gradient. Similar behaviour can be seen in some of the 
atomic indices measured in the GMOS-N spectra of that galaxy.

	CaT is more-or-less constant with radius 
in NGC 4305, matching the behaviour of Mg$_2$ and TiO$_2$. 
There are CaT gradients in both NGC 4497 and NGC 4620. 
CaT stays more-or-less constant with radius in the emission line 
galaxies NGC 4491 and NGC 4584. This behaviour differs from that defined by 
Mg$_2$ and TiO$_2$ in these galaxies, but is similar to the radial behaviour 
of the NaD index. Finally, the tendency for CaT to strengthen 
with radius in NGC 4306 is likely a consequence of 
imperfect suppression of sky emission lines, which are 
plentiful at wavelengths near the Ca triplet. 

	The CaT indices in NGC 4491 and NGC 4584 are similar to those in the other 
galaxies. This is surprising given that other metallic indices in the spectra of 
these galaxies are veiled by a young stellar continuum. The Ca triplet is at redder 
wavelengths than the other features considered here, and so is less affected by a 
blue continuum. Still, the Ca lines in the central spectra of these galaxies 
likely do not originate in the RGB and AGB stars that dominate the spectra of 
the other galaxies. Instead, the Ca triplet lines near the centers of 
NGC 4491 and NGC 4584 likely reflect a significant contribution from RSGs, which 
will be present if there has been star-forming activity during the past few 
tens of Myr. Models suggest that the equivalent width of the Ca triplet 
in populations with an age 10 Myr may be 2 - 3\AA\ higher than in populations 
with ages 100+ Myr (e.g. Figure 97 of Leitherer et al. 1999), and 
veiling by a featureless continuum will weaken these lines. 
A fortuitous amount of veiling is then required to produce CaT indices that match 
those in the galaxies with no line emission.

\subsubsection{Indices that Probe Age: H$\alpha$ and H$\beta$}

	There is a tendency for H$\beta$ to 
weaken towards larger radii in NGC 4305, NGC 4306, NGC 4497, and NGC 4620. 
In contrast, H$\alpha$ is either constant or changes only slightly with radius in these 
galaxies. Despite this seeming discrepancy between the behaviour of H$\beta$ and 
H$\alpha$, which could occur if H$\alpha$ is partly filled 
by line emission in the central few arcsec of these galaxies, 
ages estimated from H$\beta$ and H$\alpha$ are in reasonable 
agreement (Section 6), arguing that emission does not skew the H$\alpha$ measurements 
in these four galaxies. As for NGC 4491 and NGC 4584, the H$\beta$ indices of 
these galaxies {\it are} skewed by line emission at small radii, and so do not track 
Balmer line absorption in the stars that make up the main body of these galaxies. 
The negative gradients in the H$\alpha$ and H$\beta$ indices in those galaxies 
track the diminishing fractional contribution made by line emission 
towards progressively larger radii. 

	The behaviour of H$\beta$ with respect to Mg$_2$ and CaT is investigated in 
the top panels of Figures 14 and 15. NGC 4491 and NGC 4584 follow trends in 
the H$\beta$ -- Mg$_2$ plane that differ from those defined by galaxies 
that do not have an emission spectrum. Still, the agreement between the emission-free 
and emission line galaxies improves when Mg$_2 > 0.1$, as line emission is 
greatly reduced in the parts of NGC 4491 and NGC 4584 that have larger Mg$_2$ 
indices (i.e. at larger radii). This agreement suggests that the spectra at 
offsets $\geq 5$ arcsec from the centers of NGC 4491 and NGC 4584, where 
light from the disk dominates over that of the bulge, 
sample regions where H$\beta$ and H$\alpha$ may be largely free of line emission, 
or are at least subject to similar levels of contamination from line emission as 
in the galaxies that do not show obvious signs of emission in their spectra.

\begin{figure*}
\figurenum{14}
\epsscale{1.0}
\plotone{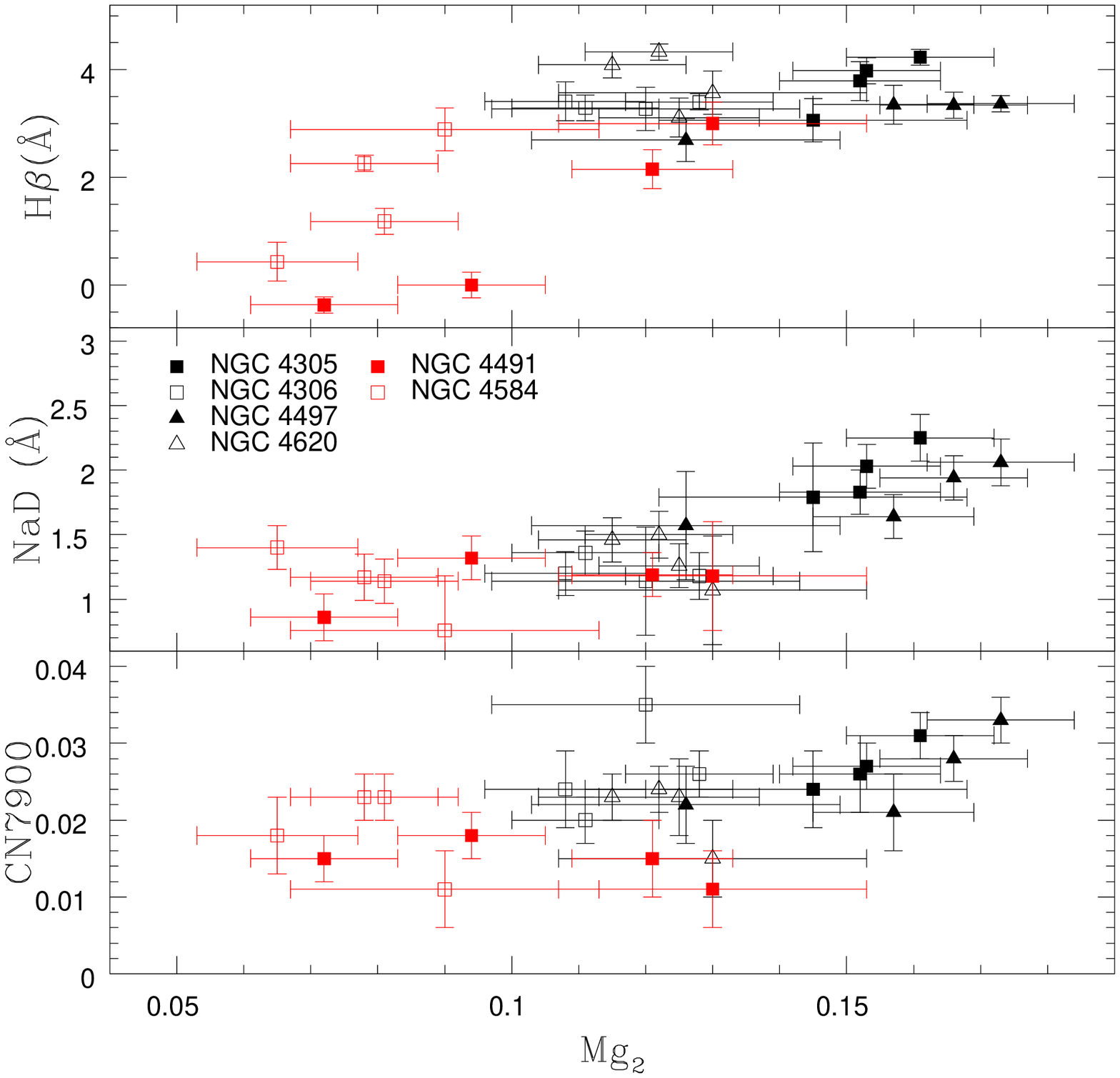}
\caption{Behaviour of H$\beta$, NaD, and CN7900 with respect to Mg$_2$. All 
measurements are from the GMOS-S spectra, and only indices measured in Regions 1 -- 4 
are shown. Points for the emission line galaxies NGC 4491 and NGC 4584 are in 
red. The H$\beta$ measurements of NGC 4491 and NGC 4584 fall along the trend defined by 
the galaxies that do not have line emission when Mg$_2 > 0.1$, hinting that the 
H$\beta$ lines of these galaxies in this Mg$_2$ regime are not significantly skewed by 
line emission.}
\end{figure*}

\begin{figure*}
\figurenum{15}
\epsscale{1.0}
\plotone{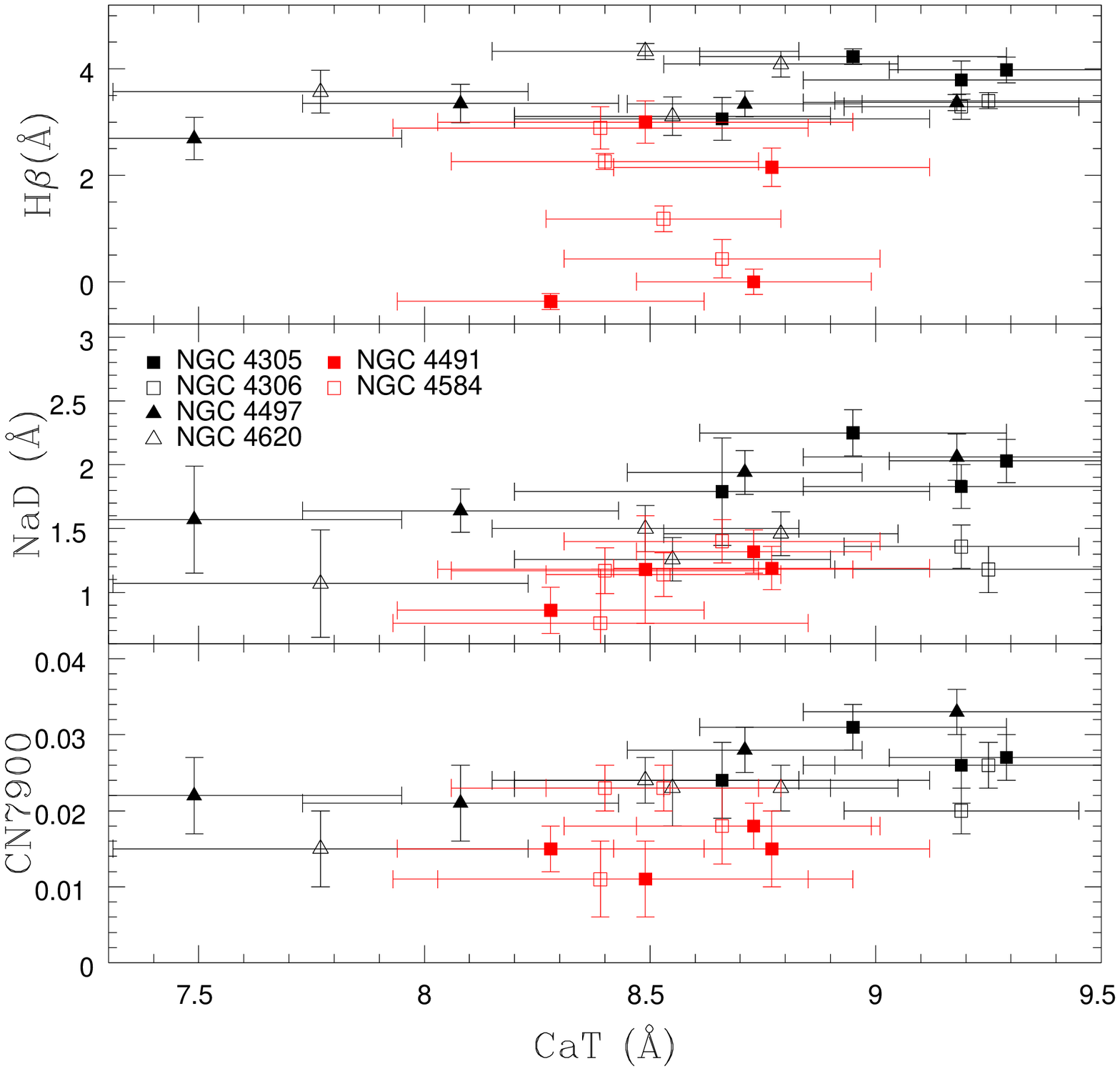}
\caption{Same as Figure 14, but with CaT along the x axis. 
The deep CaT indices for NGC 4491 and NGC 4584 at small radii are 
likely due to RSGs, which are expected to make a significant contribution to 
the integrated spectrum if there was large-scale star-forming activity during 
the past $\sim 10+$ Myr. The NaD and CN7900 indices of NGC 4491 and NGC 4584 
occupy the lower envelope of the data distributions.}
\end{figure*}

	Points from NGC 4305, NGC 4306, NGC 4497, and NGC 4620 define 
a more-or-less linear relation between H$\beta$ and CaT in the top panel of Figure 15. 
The NGC 4491 and NGC 4584 points in the upper panel of Figure 15 follow 
trends that differ from those defined by the other galaxies. It is worth noting 
that whereas most of the NGC 4491 and NGC 4584 points fall to the 
left of the main body of points in Figure 14, the CaT indices in NGC 4491 and NGC 
4584 are similar to those in the galaxies with spectra that are free of emission. 

\subsubsection{NaD}

	There is evidence for interstellar material in 
some of the target galaxies, and this could 
complicate the interpretion of the radial behaviour of NaD. 
In particular, NGC 4491 and NGC 4584 have central $J-K$ 
and $J-W2$ colors that are comparatively red, with the $J-W2$ color 
pointing to centrally-concentrated emission from hot dust (Section 3).
Barring the presence of a fortuitously large population of evolved red stars, 
these results are consistent with measureable levels of extinction 
at visible wavelengths near the centers of these galaxies. The NaD index 
is more-or-less constant with radius in both galaxies, whereas the other metallic 
features in these galaxies show a range of behaviours.
NGC 4305, NGC 4306, NGC 4497, and NGC 4620 have colors that are not indicative of 
significant dust absorption, and there is a tendency for NaD 
to weaken towards larger radii.

	The behaviour of NaD with respect to Mg$_2$ and CaT is examined in the middle 
panels of Figures 14 and 15. NaD increases more-or-less in step with Mg$_2$ 
and CaT in NGC 4305, NGC 4306, NGC 4497, and NGC 4620. There also appears to be a 
systematic offset between the NGC 4305 and NGC 4497 NaD indices at large Mg$_2$, 
in the sense that the NGC 4497 indices fall below those of NGC 4305. 
A similar offset is seen in the other panels of Figure 14, although 
a corresponding offset is not present between the NGC 4305 and NGC 4620 
sequences in the middle panel of Figure 15. This 
might point to higher [Mg/Na] in NGC 4497 when compared with NGC 4305.
La Barbera et al. (2017) find that [Na/Fe] tracks [$\alpha$/Fe] 
in the one galaxy of their sample from which they were able to extract 
spectral information over a range of radii.

	The NaD indices for NGC 4491 and NGC 4584 NaD fall along the general trend 
defined by non-emission line galaxies in Figure 14, and occupy the lower 
envelope of the NaD $vs.$ CaT distribution in the middle panel of Figure 15. There 
is overlap in the areas occupied by NGC 4491 and NGC 4584 in the middle panels of 
both figures. This overlap is of interest as the central $J-K$ and $J-W2$ colors 
of NGC 4491 and NGC 4584 differ, suggesting that there might be differences in the 
amount of obscuration and in the characteristic dust temperature. 
That NGC 4491 and NGC 4584 occupy similar areas of the NaD $vs$ Mg$_2$ and 
NaD $vs$ CaT diagrams argues that dust may not be the sole contributor to 
NaD absorption in the central regions of these galaxies.

\subsubsection{CN7900}

	CN7900 is roughly constant with radius throughout NGC 4305 and the 
inner regions of NGC 4306. Although CN7900 rises in 
the last two radial bins in NGC 4306, this is likely the result of 
uncertainties in sky subtraction. There is a tendency for CN7900 to weaken towards 
larger radius in NGC 4497 and NGC 4620. 

	The behaviour of CN7900 with respect to Mg$_2$ and CaT is 
examined in the lower panels of Figure 14 and Figure 15. The CN7900 
measurements among the four galaxies that lack line emission form a clear 
relation when Mg$_2 > 0.1$ in Figure 14. The NGC 4306 
point that falls above the other points in the lower panel of Figure 14 
is at large radii, and is likely affected by uncertainties in sky subtraction.
The otherwise tight relation defined by non-emission line galaxies in Figure 14 
suggests a small galaxy-to-galaxy dispersion in the properties of the stars that 
contribute to this feature. A corresponding relation between CN7900 
and CaT is also evident in Figure 15. The CN7900 measurements from NGC 4491 define 
the lower envelope of the data distributions 
in the lower panels of Figures 14 and 15. 

	CN7900 stays roughly constant with radius in NGC 4491 and NGC 
4584. This is a surprising result given the evidence 
for centrally concentrated young populations in these galaxies.
The CN band near 7900\AA\ is present in the 
spectra of RSGs, and these stars likely produce the deep Ca triplet lines 
seen at small radii in the spectra of these galaxies. RSGs may then be 
the source of the CN absorption in the 
spectra of NGC 4491 and NGC 4584 at small radii. 
Still, some of the CN7900 measurements in NGC 4491 and 
NGC 4584 fall along the trend defined by the other galaxies 
in the lower panel of Figure 14. Curiously, the CN7900 measurements in NGC 4491 and NGC 
4584 that are farthest from the trend defined by the other galaxies 
are at large radii, and are in the sense that CN7900 is weaker than 
in the non-emission galaxies. This hints at a possible difference in the 
stellar contents of the disks of NGC 4491 and NGC 4584 when compared 
with the disks of the other galaxies.

\section{COMPARISONS WITH MODELS}

	The ages and metallicities of the stars that dominate the integrated light 
at visible and red wavelengths have been estimated by making comparisons 
with the depths of selected features in model spectra of SSPs. While galaxies are 
complex stellar systems, comparisons with SSP models allow characteristic ages 
and metallicities of the stars that dominate the light to be determined. The results 
provide insights into the last major episode of star-forming activity 
in these galaxies. This in turn provides information for assessing their evolution 
and relationship with other galaxies in Virgo.

	Model spectra from the Bag of Stellar Tricks and 
Isochrones (BaSTI) compilation (Cordier et al. 2007; 
Percival et al. 2009) have been adopted for these comparisons. 
The BaSTI models use the evolutionary tracks described by 
Pietrinferni et al. (2004) and Cordier et al. (2007). 
These models cover a wide range of ages, metallicities, 
and wavelengths, and include evolution on the thermally-pulsing 
AGB (TP-AGB). Light from TP-AGB stars provides a progressively larger 
contribution towards longer wavelengths, and can account for a significant fraction of 
the light in the NIR. The wide wavelength coverage of these models permit consistent 
comparisons to be made with the NIR spectra that are the subject of Paper II. 
The models used here assume a Reimers (1975) mass loss coefficient of 0.4, 
and are from the high spectral resolution group of models. 

	Model spectra were downloaded from 
the BaSTI website \footnote[2]{http://basti.oa-teramo.inaf.it/}. These 
were smoothed to simulate the wavelength resolution of the 
GMOS-S spectra, and then re-sampled to match the wavelength 
cadance of the observations. The model spectra were continuum-corrected 
using the same order of fitting function as was applied to the science data.

	Comparisons are made with indices obtained from the GMOS-S spectra, as 
the depths of features in the GMOS-N spectra match those in the GMOS-S data. 
While not utilizing the full body of information that is available 
in a spectrum, spectral indices provide guidance for selecting models that 
reproduce key elements of a spectrum in a controlled manner that allows 
the information from features that are sensitive probes of age -- like H$\beta$ -- 
to be exploited. The veracity of the results and the assumptions made in 
constructing the models, such as chemical mixture, can then be 
checked by making comparisons with parts of the spectrum not used to determine 
age and metallicity.

	Comparisons with models are 
restricted to angular intervals where line emission is not detected. 
Two spectra were constructed for NGC 4305, NGC 4497, 
and NGC 4620: (1) the means of regions 1 and 2 (hereafter the `central' 
region), and (2) the means of regions 4 and 5 (hereafter the `disk'). 
Given the uncertainties in the spectra at large angular offsets due to poor 
sky subtraction then only a central spectrum was constructed for NGC 4306.
As for the galaxies with line emission, a disk spectrum was 
constructed for NGC 4491 by combining the spectra of Regions 4 and 5, whereas for NGC 
4584 -- where line emission is seen over a larger radial interval -- the 
disk spectrum is taken to be that of Region 5.

\subsection{Ages and Metallicities Estimated from Index--Index diagrams}

	H$\beta$, Mg$_2$, H$\alpha$, and CaT indices were measured from scaled-solar 
chemical mixture model spectra with metallicities Z=0.004, Z=0.008, and Z=0.019.
The results are compared with indices obtained from the galaxy spectra
in Figures 16 and 17. H$\beta$ and H$\alpha$ probe age, while 
Mg$_2$ (Figure 16) and CaT (Figure 17) are proxies for metallicity. 

	Unlike most metallic and molecular features, the depths of the 
Ca lines are relatively stable to changes in age. This is apparent in Figure 
17, where the dispersion in the CaT indices for the Z=0.008 and Z=0.019 sequences is 
$\pm 0.5 \AA$ over an age range of 0.2 Gyr to 10 Gyr. In contrast, the model 
sequences in Figure 16 slash diagonally across the H$\beta$ vs Mg$_2$ and H$\alpha$ vs. 
Mg$_2$ diagrams. Given this difference in behaviour, a comparison of the ages 
estimated from Figures 16 and 17 is of obvious interest. 

\begin{figure*}
\figurenum{16}
\epsscale{1.0}
\plotone{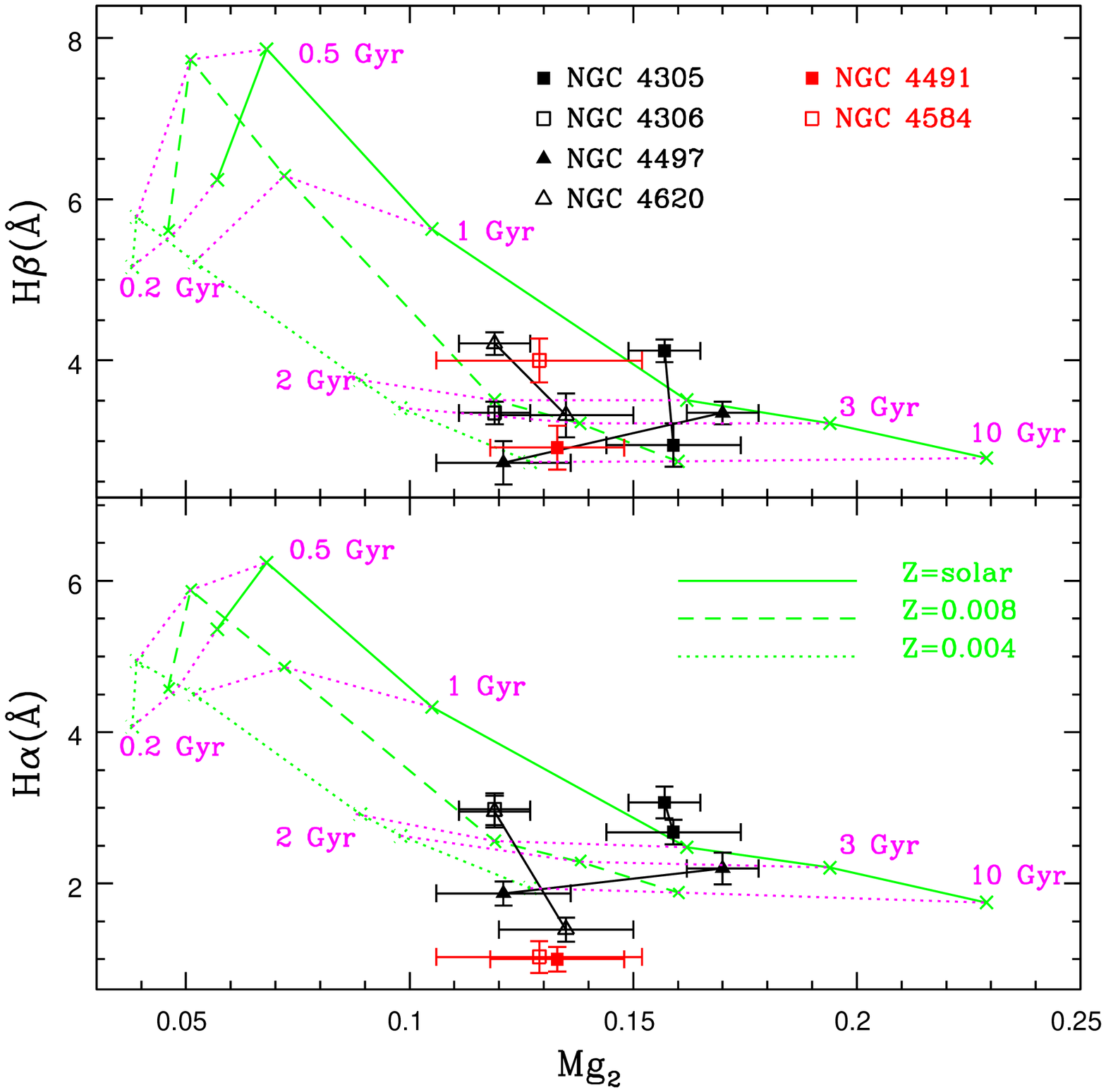}
\caption{Comparisons with indices measured from BaSTI model spectra that have been 
processed to match the characteristics of the GMOS-S spectra. Sequences constructed 
from models with ages of 0.2, 0.5, 1.0, 2.0, 3.0, and 10 Gyr are shown. The green 
sequences show model tracks for Z=0.019 (solid line), Z=0.008 (dashed line), 
and Z=0.004 (dotted line). The dotted magenta lines are isochrones. 
Points for both the centers and inner disks of NGC 4305, NGC 4497, and NGC 4620 are 
plotted, and these are joined by a line. Only points measured from the center 
of NGC 4306 and the inner disk of NGC 4491 and NGC 4584 are shown. 
The models consistently predict ages in excess of 1.5 Gyr for the centers, and the 
difference in age between the centers and inner disks of NGC 4305, NGC 4497, 
and NGC 4620 may amount to many Gyr. The H$\beta$ and H$\alpha$ indices 
tend to yield similar ages and metallicities among the 
galaxies that do not show line emission, suggesting that 
line emission does not lurk in the H$\alpha$ lines of those galaxies. 
In contrast, the H$\alpha$ indices for NGC 4491 and NGC 4584 fall 
below the model tracks, suggesting that line emission may still be present in 
the spectra of the outer regions of these systems, even though evidence of line 
emission at other wavelengths is not detected. With the obvious exception 
of the central regions of NGC 4491 and NGC 4584, the comparisons in this figure suggest 
that large-scale star formation in these galaxies ceased 1.5 Gyr or more in the past.}
\end{figure*}

\begin{figure*}
\figurenum{17}
\epsscale{1.0}
\plotone{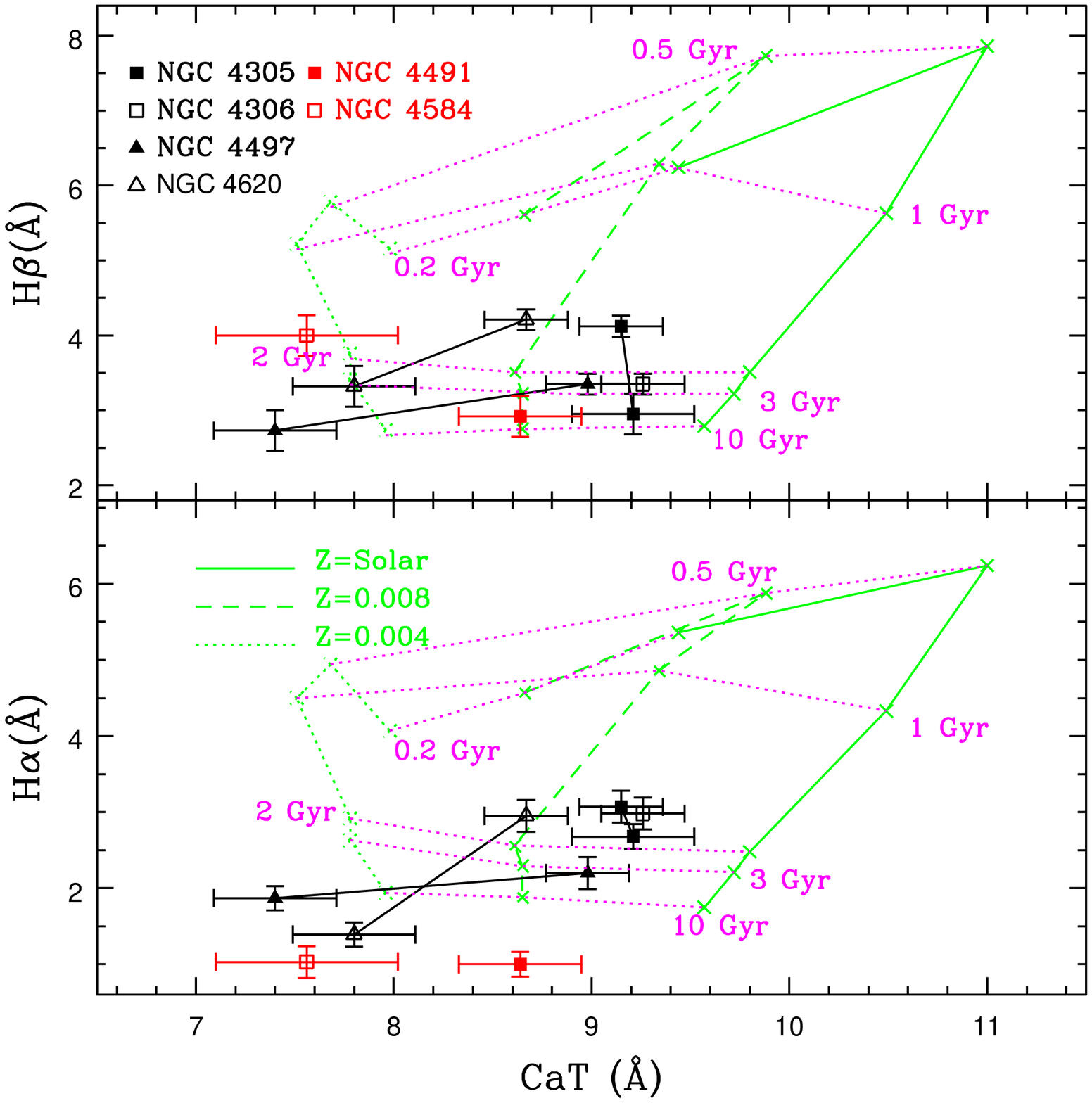}
\caption{Same as Figure 16, but with CaT along the x axis. 
Comparisons with the mean metallicities found in Figure 16 indicate 
differences of up to $\sim 0.2 - 0.3$ dex.}
\end{figure*}

	The location of the galaxies in Figures 16 and 17 indicate that there is 
only a modest age dispersion among the central regions 
of NGC 4305, NGC 4306, NGC 4497, and NGC 4620. The central regions 
of NGC 4305, NGC 4306, and NGC 4620 have luminosity-weighted ages of $\sim 1.7 - 1.9$ 
Gyr, while the center of NGC 4497 may be a Gyr older. 
The ages estimated from H$\beta$ and H$\alpha$ for 
these galaxies are in reasonable agreement. This consistency 
lends confidence to the age estimates, while also suggesting that 
line emission has not skewed the equivalent width of H$\alpha$. 
The modest dispersion in the nuclear ages of these galaxies 
hints that an external factor that is common to all 
four systems -- such as environment -- likely played a key role in 
defining the recent evolution of these systems. If this is the case then these 
galaxies are not recent additions to the Virgo cluster, but have evolved 
under its influence for at least $\sim 2$ Gyr, or at least one cluster crossing 
time (Boselli \& Gavazzi 2006).

	The inner disk points fall in parts of Figures 16 
and 17 where a small change in the equivalent 
width of H$\beta$ and H$\alpha$ can alter the age by a Gyr or more, and so 
the age dispersion among the inner disks of these galaxies may be 
larger than between the central regions. The outer 
regions of NGC 4305, NGC 4497, and NGC 4620 have ages that are older 
than the central regions of their hosts, although the H$\alpha$ measurements of 
NGC 4305 are consistent with a modest difference in age between the central and 
inner disk of that galaxy. The age of the inner disk of NGC 4620 estimated from 
H$\beta$ is $\sim 4$ Gyr, whereas an age $> 10$ Gyr is estimated from 
H$\alpha$. This suggests that emission may partially fill H$\alpha$ in the disk 
spectrum of that galaxy. The H$\beta$ and H$\alpha$ measurements for 
the inner disk of NGC 4497 are both consistent with an age of 10 Gyr. NGC 4497 
also has the oldest central regions, with an age $\sim 3$ Gyr. 
NGC 4497 thus appears to be the galaxy in our sample for which star formation has 
been truncated the longest.

	As for NGC 4491 and NGC 4584, if H$\beta$ in 
the outer region of NGC 4584 is free of line emission then it has an age 
that is comparable to the centers of the other galaxies. The outer regions 
of NGC 4491 may be even older. It thus appears that -- with the obvious exception 
of the central regions of NGC 4491 and NGC 4584 -- 
there has not been large-scale star formation in these galaxies since intermediate 
epochs.

	Previous studies have estimated the age of NGC 4305. 
The blue/yellow spectrum of NGC 4305 was examined by 
Fraser-McKelvie et al. (2016) in a study of six candidate passive spiral galaxies. 
They find that NGC 4305 has the deepest H$\beta$ and H$\alpha$ lines in their sample. 
They estimate a characteristic age of 1 Gyr for NGC 4305 from the D4000 index, 
whereas the other galaxies in their sample have ages $\sim 1.4 - 3.2$ Gyr, with 
a mean of 2.3 Gyr. We find an age of $\sim 1.8$ Gyr 
for the central regions of NGC 4305 based on the depths of the Balmer lines.

	Roediger et al. (2012) used broad-band colors extracted from $griH$ 
observations to estimate the age of NGC 4305, and found an age of $\sim 10+$ Gyr. 
The GMOS spectra predict a much younger age for NGC 4305 than that found by 
Roediger et al. (2012). There is considerable radial jitter in the photometric age 
estimates, but when averaged over larger angular 
scales there is an overall tendency for the Roediger et al. age estimates to 
stay constant or increase with increasing radius. The radial behaviour of 
H$\alpha$ and H$\beta$ in the GMOS spectra is consistent with an age gradient.

	The metallicities inferred from the models in Figures 16 and 
17 differ from galaxy-to-galaxy. The central regions of NGC 4305 and NGC 
4497 are the most metal-rich in the sample, and the Mg$_2$ indices are indicative of 
solar or possibly even slightly super-solar metallicities for 
the centers of these galaxies. However, the CaT indices 
indicate metallicities that are $\sim 0.2 - 0.3$ dex lower. 

\begin{figure*}
\figurenum{18}
\epsscale{1.0}
\plotone{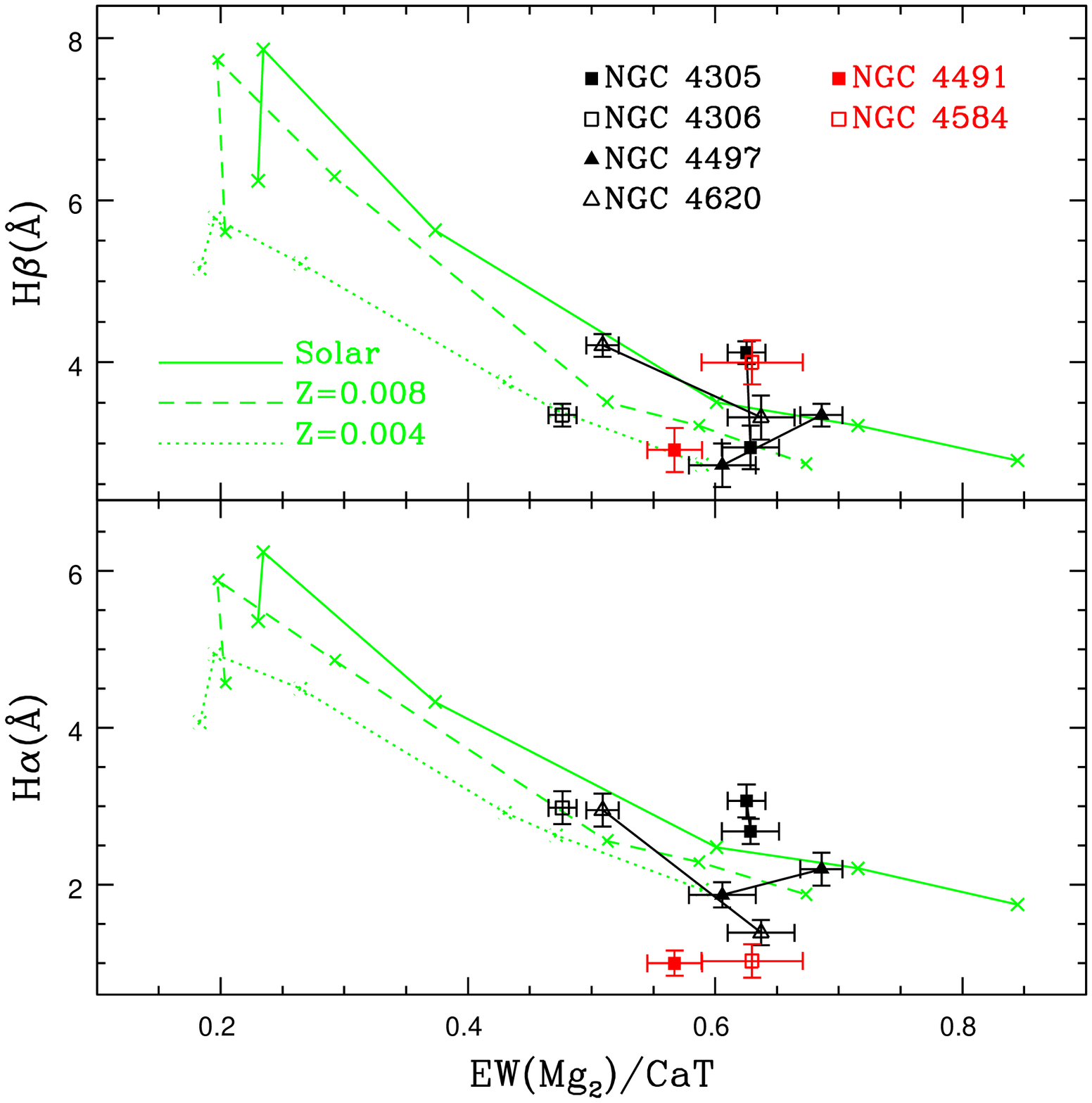}
\caption{Ratio of the equivalent 
widths of Mg$_2$, calculated from the Mg$_2$ magnitude measurements, and CaT. 
The observed points in the top panel fall along the model sequences, 
while some galaxies deviate from the model sequences in the lower panel. 
The locations of NGC 4305, NGC 4306, and NGC 4497 on these diagrams is more-or-less 
consistent with their locations on Figure 16, signaling agreement with 
the scaled-solar chemical mix adopted for the models.}
\end{figure*}

	To further examine if [Mg/Ca] differs from that assumed by the 
models, the behaviour of H$\beta$ and H$\alpha$ with respect to the ratio of Mg$_2$ to 
CaT are shown in Figure 18. The equivalent width of Mg$_2$ in \AA\ was calculated 
from the Mg$_2$ magnitude measurements. The Mg$_2$/CaT ratio 
changes with metallicity due to the overall shift in color of the evolutionary 
sequences and the resulting changes in the contribution that bright main sequence stars 
make to the light at wavelengths near Mg$_2$.

	The galaxies scatter about the model 
predictions in the top panel of Figure 18. Some galaxies depart from the model 
sequences in the bottom panel, and this is attributed to contamination from H$\alpha$ 
emission. The locations of NGC 4305, NGC 4306, and NGC 4497 in Figure 18 
are consistent with their locations in Figure 16, suggesting that the 
[Mg/Ca] mix is consistent with that assumed for the models.

\subsection{Comparisons with Fe Lines and NaD}

	The model spectra match the depths of other features not considered in 
Figures 16 and 17, and this is demonstrated in Figure 19 where 
comparisons are made with the central spectra of NGC 4305, NGC 4306, NGC 4497, 
and NGC 4620. Spectra of NGC 4491 and NGC 4584 are not shown given that continuum 
emission will veil absorption features in their central spectra. 
The wavelength regions shown in this figure contain deep Fe 
lines (left hand panel), and the NaD blend (right hand panel). Comparisons are made 
with Z=0.008 and Z=0.019 models with an age of 2 Gyr. This age and range of 
metallicities reflects those found from Figures 16 and 17. 

\begin{figure*}
\figurenum{19}
\epsscale{1.0}
\plotone{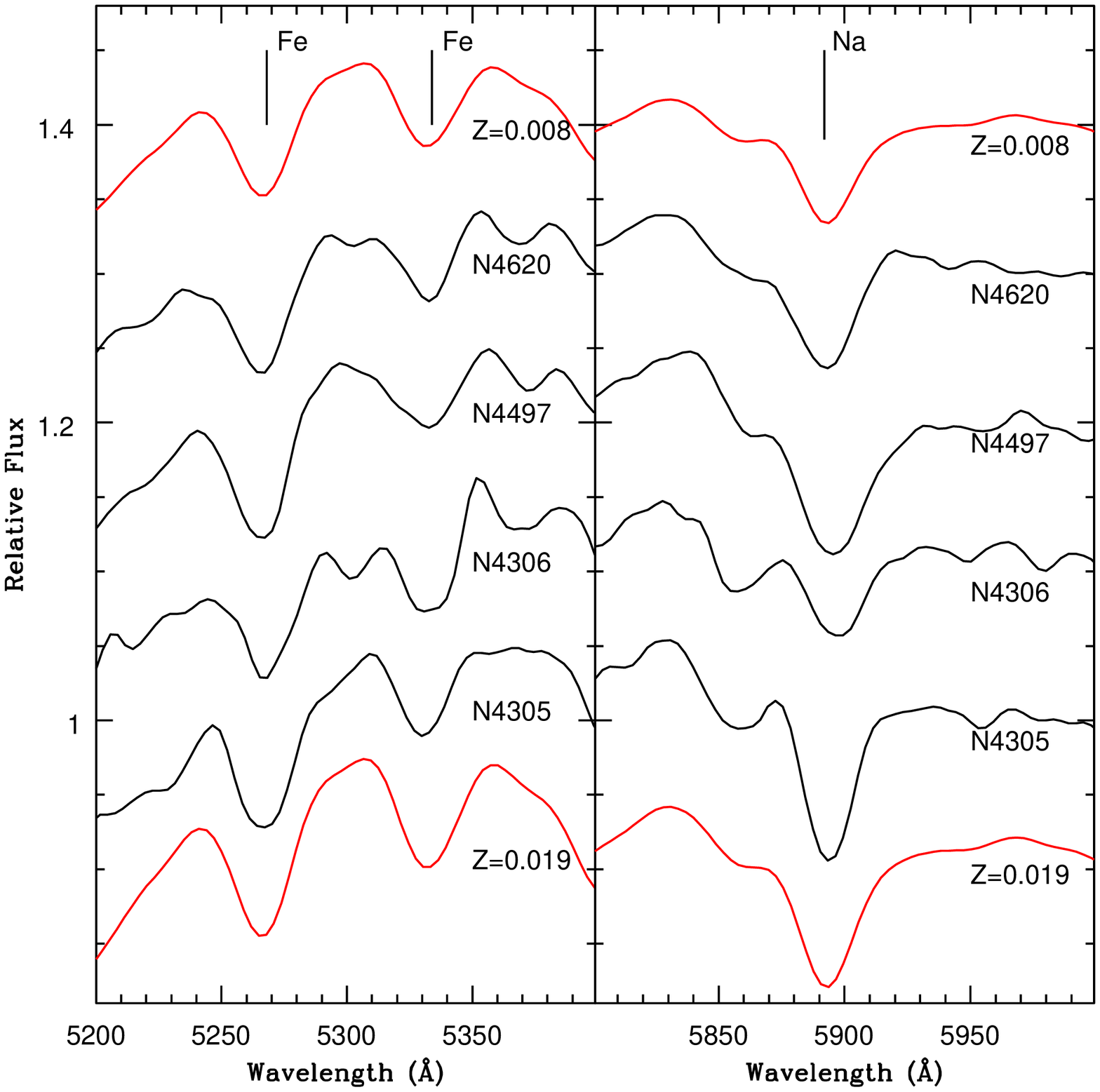}
\caption{Wavelength intervals that sample Fe and NaD lines in the central 
spectra of the four galaxies with no obvious line emission. 
Models with ages 2 Gyr and metallicities Z=0.008 and Z=0.019 
are shown as red lines. The Fe features for NGC 4305 are more-or-less 
consistent with Z=0.019, while NaD is deeper than predicted from 
the models. NGC 4305 is unique in this respect, and this could suggest that 
[Na/Fe] $> 0$ in this galaxy. The Fe lines in the NGC 4306 spectrum are matched 
by the Z=0.008 model, while the NaD line is weaker than predicted by the Z=0.008 model. 
The Fe and NaD features in the NGC 4497 spectum are matched by the Z=0.019 model, 
while these lines in the NGC 4620 spectrum have depths that match 
those in the Z=0.008 model.}
\end{figure*}

	The Fe lines near 5270\AA\ and 5335\AA\ are among 
the strongest Fe features in the visible wavelength region. 
The Fe lines provide insights into chemical mixture and total metallicity 
when compared with features like the Mg triplet near 5170\AA\ . The depths 
of the Fe lines at 5270 and 5335\AA\ in Figure 19 generally fall within 
the range defined by the Z=0.008 and Z=0.019 models. The deepest Fe lines 
are in NGC 4305, and these have a depth that is consistent with Z=0.019. The weakest 
Fe lines are in NGC 4306, which are matched best 
by the Z=0.008 model. The relative depths of the Fe lines in these two galaxies 
is thus consistent with their locations in Figure 16. 
In contrast, the positions of NGC 4305 and NGC 4306 in Figure 17 suggests that 
the centers of these galaxies have similar metallicities. This 
is not consistent with the depths of the Fe lines in the spectra of these galaxies.
As for the other galaxies, the Fe lines in NGC 4497 are intermediate in 
strength between the Z=0.008 and Z=0.019 models, while those in the NGC 4620 spectrum 
tend to match the Z=0.008 models. Thus, the relative strengths of the Fe lines in 
NGC 4497 and NGC 4620 are more-or-less consistent with the metallicities inferred from 
Figures 16 and 17.

	Of the various Na features at visible, red, and NIR 
wavelengths, NaD is the most sensitive to [Na/Fe] (La Barbera et al. 2017). 
While the depth of NaD can be affected by interstellar absorption, the colors 
of the galaxies examined in Figure 19 suggest that there are not large amounts of 
absorbing material, and so the depth of NaD either reflects [Na/Fe] or a mass function 
that contains a different fraction of low mass stars than considered in the models. 
We consider the latter possibility to be unlikely given the depths of other features 
in the spectrum, and this issue will be revisited in the next paper of this series 
(Davidge 2018, in preparation). Inspection of Figure 19 shows that the depth of 
NaD in the central spectrum of NGC 4305 is consistent with Z $> 0.019$, 
and NGC 4305 is the only galaxy to show stronger than expected NaD absorption. NaD is 
shallowest in NGC 4306, where it is consistent with -- or even weaker than -- 
the Z=0.008 model. To the extent that NGC 4305 and NGC 4306 have a 
solar neighborhood-like mass function then [Na/Fe] $\neq 0$
in their central regions. An enhanced [Na/Fe] has been 
found in other early-type galaxies (La Barbera et al. 2017).

	In summary, the behaviour of features that were not considered when 
estimating age and metallicity in Figures 16 and 17 
tend to be consistent with the results found from 
Figure 16, where Mg$_2$ is a proxy for overall metallicity, rather 
than Figure 17, where CaT is the proxy. 
Thomas et al. (2003a) argue that Ca may be underabundant in 
early-type galaxies, and this is one possible explanation for some of the 
differences between the metallicities deduced from 
Mg$_2$ and CaT in Figures 16 and 17. Nevertheless, 
that there appears to be a near-solar Mg and Fe mixture argues that 
these galaxies experienced prolonged chemical enrichment before star formation 
was truncated. 

\section{DISCUSSION AND SUMMARY}

	Deep long slit spectra have been used to investigate 
the stellar contents of six moderate mass early-type disk galaxies in 
the Virgo cluster. Five of the galaxies were classified by Lisker et al. (2006a) 
as 'dwarf-like' S0/Sa. These objects differ from their more 
massive counterparts in that their central light profiles 
are more akin to those of dEs, rather than the bulge-dominated central regions 
of classical lenticulars. The disks of two galaxies (NGC 4305 and NGC 4620) 
have noticeable spiral structure. The sixth galaxy -- NGC 4306 -- is 
not classified as a dwarf-like S0/Sa. However, it is still related to the others, 
as Lisker et al. (2006a) found it to be one of the few Virgo dE galaxies 
to have a `certain disk'. The galaxies have similar integrated K-band magnitudes, 
indicating similar total masses. Their projected location on the sky suggests that 
they sample a range of environments within the cluster. 

	The stellar contents of these galaxies provide insights into 
the nature of their progenitors and the mechanisms that have shaped 
their present-day properties. To this end, comparisons are made 
with model spectra to estimate luminosity-weighted ages and metallicities. 
A key result is that the dominant source of light from the central regions of all six 
galaxies is not an old population, but instead is either a young or an intermediate 
age population. There is a factor of two dispersion in the 
metallicities of NGC 4305, NGC 4306, NGC 4497, and NGC4620, 
and the luminosity-weighted ages of their central regions are 
$1.7 - 3$ Gyr, indicating that there has not been recent large-scale star formation. 
The characteristic ages for the centers of these four galaxies are on the same 
order as the crossing time of the Virgo cluster (1.7 Gyr; Boselli \& Gavazzi 2006), and 
overlap with the truncation times for large-scale star formation in a sample of early 
and late-type galaxies in Virgo selected from the HRS (Boselli et al. 2016).
Ages for these galaxies estimated from H$\alpha$ and 
H$\beta$ tend to agree, indicating that significant 
emission is not lurking in the H$\alpha$ line. Some of the galaxies 
harbor radial age gradients, in the sense of older ages at larger 
radii. This is consistent with the radial behaviour of broad-band radial colors 
of dE(is)s by Lisker et al. (2008).

	There are caveats to the ages estimated here. 
The spectrograph slit samples only a modest fraction of each 
galaxy. However, any stars that formed during recent large-scale star formation would 
disperse throughout the galaxy over the course of a few disk rotation times (i.e. 
within the past few hundred Myr), and so should be detected if they occured in large 
numbers. Another caveat is that luminosity-weighted ages depend on wavelength, and 
observations at other wavelengths may recover additional insights into these systems. 
Even modestly-sized young populations can dominate the SED in the UV 
while leaving no signatures at visible wavelengths (e.g. Vazdekis et al. 2016). 
Of the four galaxies with intermediate-aged central regions, NGC 4497 is the only one  
that was imaged with GALEX. The center of NGC 4497 is barely detected 
in the FUV channel. If a young component is present near the center of 
NGC 4497, it makes only a puny contribution to the overall central mass. This is 
consistent with the 3--4 Gyr age estimated for the central regions of this galaxy. 

	NGC 4491 and NGC 4584 differ from the other galaxies in that there 
are signs of very recent star-forming activity in their central regions. 
The relative strengths of [SII]6746 and H$\alpha$ emission in the spectra 
of both galaxies is consistent with photoionization by massive hot stars. 
The comparatively red central $J-W2$ colors of both galaxies further suggests that 
centrally-concentrated hot dust is present, which could be heated by 
massive, hot main sequence stars. Neither galaxy was flagged by Lisker et 
al. (2006b) as having a blue nucleus or a morphology similar to that of blue 
nucleated dEs. This non-detection is likely due to dust obscuration that reddens 
the star-forming region, and highlights the importance of spectra 
for identifying young populations in early type dwarf galaxies in Virgo.

	Examining the mechanism(s) that triggered the recent star-forming 
activity in NGC 4491 and NGC 4584 might provide insights 
into the evolution of the other four galaxies. The angular offsets from M87 listed in 
Table 1 indicate that both NGC 4491 and NGC 4584 have projected distances of only 
a few hundred kpc from this galaxy, which lies near the center of 
the Virgo cluster. If these galaxies retained external dust reservoirs within 
the past $\sim 1$ Gyr, then interactions with other galaxies may 
have sparked galaxy-wide levels of elevated star formation, and caused gas and dust 
to be funneled into the inner regions of these galaxies. 
Elevated levels of star-forming activity in star bursts may last a few hundred 
Myr (e.g. McQuinn et al. 2010), and the last vestages of this activity 
tend to be found in the central regions of a galaxy. Centrally-concentrated 
star-forming activity can remain for some time after star formation 
in the outer regions of the galaxy has been curtailed (e.g. Soto \& Martin 2010). 
However, the comparatively old ages of the disks of NGC 4491 and NGC 4584 hints that 
the present-day star-forming activity in these galaxies is not the remnant of a 
classical galaxy-wide star burst. 

	Grootes et al. (2017) suggest that spiral galaxies in dense 
environments that have been stripped of gas may be 
able to re-acquire gas by accreting material from the intrahalo medium (IHM). 
If the gas acquired from the IHM has no angular momentum, 
then it will dilute the overall angular momentum content of the host disk. 
If significant quantities of low angular momentum IHM gas can be accreted in 
the dynamically hot environment like Virgo, then it could provide material to 
fuel centrally-concentrated star formation in NGC 4491 and NGC 4584. If a significant 
fraction of the metals in the IHM were formed early-on in the evolution of Virgo then 
the accreted gas will show chemical signatures of rapid star formation, such as 
[O/Fe] $> 0$. Still, if accretion of this nature is at play in NGC 4491 and NGC 4584 
then why are elevated levels of star formation not seen in the centers of the other 
galaxies? Also, if gas is accreted and retained then there might be spatially 
extended low ionization emission (e.g. Belfiore et al. 2017), and evidence for 
this is not seen.

	While NGC 4491 and NGC 4584 share evidence for recent star formation, it does 
not mean that the same mechanism for fueling and triggering star formation is at work 
in both galaxies. In fact, the central structural properties of these galaxies are 
very different. The Sersic index measured for NGC 4491 by McDonald et al. (2011) 
is at the low end of that found in late-type spirals and dEs. This measurement is 
likely influenced by the bar in NGC 4491, and such a structure may trigger 
the infall of star-forming material into the center of that galaxy. 
In contrast, McDonald et al. (2011) find that a Sersic index of 
1.3 matches the light profile of the central regions of NGC 
4584 in $H$. Such a Sersic index is similar to those measured from 
the central light profiles of classical Sa and Sb galaxies in Virgo (Table 2 
of McDonald et al. 2011), and suggests that NGC 4584 may contain a classical bulge. 
If NGC 4584 contains a classical bulge then the star-forming activity in 
its central regions may be independent of events at larger radii.
Indeed, Morelli et al. (2016) find that the properties of bulges in isolated 
galaxies are not related to overall morphology, and argue that bulges and 
disks evolve separately.

	While there is a wide dispersion in the central spectrophotometric properties 
of all six galaxies, there is only a modest galaxy-to-galaxy dispersion in the depth 
of H$\beta$ and Mg$_2$ a few arcsec from the galaxy centers. This agreement 
extends to broad band colors, as the $J-K$ colors of all six galaxies 
fall between 0.9 and 0.95 at major axis distances of 10 arcsec from the nucleus. 
The galaxy-to-galaxy agreement in the spectrophotometric properties 
of the disks is perhaps surprising, given that some of the galaxies have 
spiral structure. NGC 4305 hosts the most prominent spiral arms (Lisker 
et al. 2006a), but the H$\beta$ index of its disk is not 
indicative of a markedly younger luminosity-weighted age than in the other galaxies. 
That the galaxy disks have similar old ages suggests that these galaxies 
have been in Virgo for some time, or that they were anemic prior to accretion 
onto the cluster. The similarity in inner 
disk ages despite a broad range in central ages is consistent 
with the the evolution of the nuclei being defined by the characteristics of the 
nucleus itself, as suggested by C\^{o}t\'{e} et al. (2006). 

	Despite evidence that these galaxies are not recent additions to 
the Virgo cluster and signs that their interstellar mediums have been depleted,
there are indications that these galaxies have not been 
subject to large-scale stripping of their stars, 
and that they have had sustained star-forming periods on Gyr timescales 
prior to the removal of gas. In regard to the absence of large-scale stellar 
stripping, the model spectra that best match those of 
the galaxy centers and inner disks have metallicities that are consistent with 
their integrated brightnesses. That the galaxies do not deviate significantly from 
the luminosity--metallicity relation suggests that they were not stripped of 
large amounts of stars or interstellar material during their early evolution, 
when the bulk of their stars presumably formed. 

	Evidence for extended star-forming episodes comes from the chemical 
mixture. Stars in a system that has experienced more-or-less continuous (at least when 
averaged over the timescale for the onset of SNeII) star-forming activity will have 
chemical mixtures that differ from those of stars in a system that has had 
sporatic star-forming events interupted by the purging of star-forming material. 
In Section 6 it was shown that the metallicities estimated from the Mg$_2$ 
and CaT indices do not always agree, in the sense that the 
metallicities found using the CaT index may be lower than those 
based on Mg$_2$. Some of the galaxies may then have [Ca/Mg] that 
differs from the solar mixture assumed for the models. Still, the depths of Fe 
features tend to be consistent with the metallicity deduced from Mg$_2$, 
rather than the CaT. Given the difference in time spans in the production of 
Mg and Fe then this suggests that star-forming material must have been retained and 
enriched for some time in these galaxies.

	Based on the diffuse spatial distribution of disky dwarf galaxies in 
Virgo, Lisker et al. (2006a) suggest that the orbits of dE(di)s have not yet been 
virialized, implying that they may be more recent additions to Virgo 
than the galaxies that define concentrated spatial distributions. 
Lisker et al. (2006b; 2007) argue that dE(di)s were originally late-type spiral 
galaxies that have been subjected to harassment by Virgo members. Toloba et al. 
(2012) examine the structural properties of a sample of bright dEs in Virgo, 
half of which are dE(di)s, and the structural characteristics of these galaxies are 
consistent with them being transformed late-type galaxies. A similar conclusion was 
reached by Toloba et al. (2015) using a larger sample of dEs. Toloba et al. (2015) 
argue that the galaxies in their sample experienced gas removal on rapid timescales 
that are more in line with ram pressure stripping, rather than harassment. However, the 
Gyr or more difference in age between the central regions of these galaxies 
and their disks exceeds the time scale expected for ram pressure stripping in a
dense environment (e.g. Steinhauser, Schindler, \& Springel 2016), suggesting 
that ram pressure stripping alone can not explain the SFHs of these galaxies.

	NGC 4306, NGC 4491, and NGC 4497 contain bars, and 
these structures may induce radial mixing of stellar content. 
In the case of  NGC 4491, the young centrally-concentrated 
population in this galaxy complicates efforts to assess any abundance gradient. 
While the radial coverage of NGC 4306 is limited, there is no evidence of an 
abundance gradient in its central regions (Section 5). The depths of 
metallic lines vary with radius in NGC 4497 (Section 5), due to 
a radial variation in mean metallicity (Section 6). Thus, there is no 
evidence that the bar in NGC 4497 has mixed the stellar content of this galaxy.

	We close the discussion by noting that while 
the properties of these galaxies have likely been affected by 
environment, secular processes may also have played a role in shaping their 
present-day properties. Kwak et al. (2017) examines the evolution of a simulated 
low-mass disk galaxy that is based on the 
dE(di) VCC 856, which has an M$_B$ that is $\sim 0.5$ magnitudes 
fainter than NGC 4306. When evolved in isolation, Kwak et al. (2017) find that 
a bar forms at two different times, and the eventual buckling 
of these bars produces peanut and X-shaped bulges. Therefore, while some of the 
galaxies studied here have bars, these may not be the result of interactions. 
Spiral structure also forms from material that originates at the bar ends
in these simulations; the spiral structure seen in the present-day galaxies 
likely does not reflect the original morphology of the progenitor.

\acknowledgements{Thanks are extended to the anonymous referee for comments that 
helped improve the paper.}

\end{document}